\documentclass[10pt,conference,a4paper]{IEEEtran}

\IEEEoverridecommandlockouts

\usepackage{etex}
\usepackage[usenames,dvipsnames]{xcolor}
\usepackage{newlfont}
\usepackage{enumitem}
\usepackage{graphicx,float} 
\usepackage{microtype} 
\usepackage{placeins} 
\usepackage{tikz}
\usepackage{amsmath}
\usepackage{stmaryrd}
\usepackage{amsfonts}
\usepackage{amssymb,amsthm,latexsym}
\usepackage{turnstile}
\usepackage{mathrsfs}
\usepackage{calc}
\usepackage[capitalise]{cleveref}
\usepackage{times}
\usepackage{listings}
\usepackage{wrapfig}
\usepackage[arrow,matrix,frame,curve,cmtip]{xy}\CompileMatrices

\usetikzlibrary{shapes,arrows} 
\usetikzlibrary{positioning}
\usetikzlibrary{chains}
\usetikzlibrary{automata}
\usetikzlibrary{patterns}
\usetikzlibrary{decorations.pathreplacing}
\usetikzlibrary{calc,decorations.markings,decorations.pathreplacing,fit,backgrounds,shapes.symbols,shapes.geometric}
\tikzset{
  gnode/.style={draw,shape=circle,inner sep=0,minimum height=.2cm,
    minimum width=.2cm},
  hyperedge/.style={shape=rectangle,draw,inner sep=0,minimum width=.6cm,
    minimum height=.4cm}
}

\tikzset{every fit/.style={shape=rectangle,inner sep=5pt}}

\usepackage[colorinlistoftodos, textsize=footnotesize,color=orange!70]{todonotes}

\usepackage[T1]{fontenc}

\usepackage{algorithm}
\usepackage{algpseudocode}

\usepackage[sort,compress,noadjust]{cite}

\newcommand{\full}[1]{#1}
\newcommand{\short}[1]{}

\renewcommand{\epsilon}{\varepsilon}

\newcommand{\irreducibles}{\mathcal J(\mathbb L)}

\theoremstyle{plain}
\newtheorem{theorem}{Theorem}
\newtheorem{proposition}{Proposition}
\newtheorem{lemma}{Lemma}
\newtheorem{defn}{Definition}
\newtheorem{example}{Example}
\theoremstyle{definition}
\newtheorem{myalgorithm}{Algorithm}
\newcommand{\history}[1]{\downarrow\! {#1}}

\newcommand{\app}[1]{\scalebox{0.9}{$\lfloor$} #1\scalebox{0.9}{$\rfloor$}}
\newcommand{\approxL}[1]{\scalebox{0.9}{$\lfloor\!\!\lfloor$}
      #1 \scalebox{0.9}{$\rfloor\!\!\rfloor$}}

\usepackage{ifthen}
\usepackage{rotating}

\renewcommand{\phi}{\varphi}
\renewcommand{\epsilon}{\varepsilon}

\author{\IEEEauthorblockN{Harsh Beohar}
  \IEEEauthorblockA{Universit\"at Duisburg-Essen} \and
  \IEEEauthorblockN{Barbara K\"onig} \IEEEauthorblockA{Universit\"at
    Duisburg-Essen} \and \IEEEauthorblockN{Sebastian K\"upper}
  \IEEEauthorblockA{Universit\"at Duisburg-Essen} \and
  \IEEEauthorblockN{Alexandra Silva} \IEEEauthorblockA{University
    College London}}

\title{Conditional Transition Systems with Upgrades\thanks{Research
    partially supported by DFG project BEMEGA and ERC Starting Grant
    ProFoundNet (grant agreement 679127).}}

\IEEEoverridecommandlockouts\IEEEpubid{\makebox[\columnwidth]{978-1-5386-1925-4/17/\$31.00~\copyright~2017 IEEE \hfill} \hspace{\columnsep}\makebox[\columnwidth]{ }}

\begin{document}

\maketitle

\vspace{-0.4cm}

\begin{abstract}
We introduce a variant of transition systems, where activation of
transitions depends on conditions of the environment and upgrades during
runtime potentially create additional transitions. Using a cornerstone
result in lattice theory, we show that such transition systems can be
modelled in two ways: as conditional transition systems (CTS) with a
partial order on conditions, or as lattice transition systems (LaTS),
where transitions are labelled with the elements from a distributive
lattice. We define equivalent notions of bisimilarity for both variants
and characterise them via a bisimulation game.

We explain how conditional transition systems are related to featured
transition systems for the modelling of software product
lines. Furthermore, we show how to compute bisimilarity symbolically via
BDDs by defining an operation on BDDs that approximates an element of a
Boolean algebra into a lattice. We have implemented our procedure and
provide runtime results.\end{abstract}

\section{Introduction}
\label{sec:intro}

Conditional transition systems (CTS) have been introduced in
\cite{ABHKMS12} as a model for systems whose behaviour is guarded by
different
conditions. 
Before an execution, a condition is chosen by the environment from a
pre-defined set of conditions and, accordingly, the CTS is
instantiated to a classical labelled transition system (LTS).  In this
work, we consider \emph{ordered} sets of conditions which allow for a
change of conditions during runtime. It is allowed to replace a
condition by a smaller condition, called upgrade. An upgrade activates
additional transitions compared to the previous instantiation of the
system.

Our focus lies on formulating a notion of behavioural equivalence,
called \emph{conditional bisimilarity}, that is insensitive to changes
in behaviour that may occur due to upgrades. Given two states, we want
to determine under which conditions they are behaviourally
equivalent. To compute this, we adopt a dual, but equivalent, view
from lattice theory due to Birkhoff to represent a CTS by a lattice
transition system (LaTS). In general, LaTSs are more compact in nature
than their CTS counterparts. Moreover, we also develop an efficient
procedure based on matrix multiplication to compute conditional
bisimilarity.

Such questions are relevant when we compare a system with its
specification or we want to modify a system in such a way that its
observable behaviour is invariant. Furthermore, one requires
minimisation procedures for transition systems that are potentially
very large and need to be made more compact to be effectively used in
analysis.


An application of CTSs with upgrades is to model systems that deteriorate over time. Consider a system that is dependent on components that break over time or require calibration, in particular sensor components. In such systems, due to inconsistent sensory data from a sensor losing its calibration, additional behaviour in a system may be enabled (which can be modelled as an upgrade) and chosen nondeterministically.

Another field of interest, which will be explored in more detail, are software product lines (SPLs). SPLs refer to a software engineering method for managing and developing
a collection of similar software systems with common features.
To ensure correctness of such systems in an efficient way, it is
common to specify the behaviour of many products in a single
transition system and provide suitable analysis methods based on
model-checking or behavioural equivalences (see \short{\cite{DBLP:conf/icse/CordyCPSHL12,terBeek2016:MTS,Classen:2013:FTS,Atlee:2015:MBI:2820126.2820133,Classen:2010:MCL:1806799.1806850,Classen:2011:symbolic,Dubslaff:2014:PMC,Chrszon2016:profeat}}\full{\cite{DBLP:conf/icse/CordyCPSHL12,terBeek2016:MTS,Classen:2013:FTS,Atlee:2015:MBI:2820126.2820133,Classen:2010:MCL:1806799.1806850,Classen:2011:symbolic,Dubslaff:2014:PMC,Chrszon2016:profeat,Gruler:2008:PL-ccs}}).

Featured transition systems (FTS) -- a recent extension of conventional transition system proposed by Classen et al.\cite{Classen:2013:FTS} -- have become the standard formalism to model an SPL.
An important issue usually missing in the theory of FTSs is the notion of
self-adaptivity \cite{Cordy2013:adaptivefts}, i.e., the view that
features or products are not fixed a priori, but may change during
runtime. We will show that FTSs can be considered as CTSs without upgrades where the conditions are the powerset of the features. Additionally, we propose to incorporate a notion of upgrades into software product lines, that cannot be captured by FTSs.
Furthermore, we also consider deactivation of transitions in
\short{\cite{bkks:cts-upgrades-arxiv}}\full{Appendix~\ref{sec:deactivating-transitions}},
to which our techniques can easily be adapted, though some
mathematical elegance is lost in the process.

Our contributions are as follows. First, we make the different levels of granularity -- features, products and sets of products -- in the specification of SPLs explicit and give a
theoretical foundation in terms of Boolean algebras and lattices.
Second, we present a theory of behavioural equivalences with
corresponding games and algorithms and applications to conventional and adaptive SPLs.
Third, we present our implementation based on binary decision diagrams
(BDDs), which provides a compact encoding of a propositional formula
and also show how they can be employed in a lattice-based setting.
Lastly, we show how a BDD-based matrix multiplication algorithm
provides us with an efficient way to check bisimilarity 
relative to the naive approach of checking all products separately.

This paper is organised as follows. Section~\ref{sec:preliminaries}
recalls the fundamentals of lattice theory relevant to this paper.
Then, in Section~\ref{sec:cts} we formally introduce CTSs and
conditional bisimilarity. In Section~\ref{sec:lats}, using the
Birkhoff duality, it is shown that CTSs can be represented as lattice
transition systems (LaTSs) whose transitions are labelled with the
elements from a distributive lattice. Moreover, the bisimilarity
introduced on LaTSs is shown to coincide with the conditional
bisimilarity on the corresponding CTSs. In
Section~\ref{sec:matrix-mult}, we show how bisimilarity can be
computed using a form of matrix multiplication. Section~\ref{sec:spl}
focusses on the translation between an FTS and a CTS, and moreover, a
BDD-based implementation of checking bisimilarity is laid out. Lastly,
we conclude with a discussion on related work and future work in
Section~\ref{sec:conclusion}. All the proofs can be found in
\short{\cite{bkks:cts-upgrades-arxiv}}\full{Appendix~\ref{sec:proofs}}.

\section{Preliminaries}
\label{sec:preliminaries}

We now recall some basic definitions concerning lattices, including
the well-known Birkhoff's duality result from \cite{dp:lattices-order}.

%
\begin{defn}[\rm Lattice, Heyting Algebra, Boolean Algebra]
  \label{def:lattice}
  Let $(\mathbb{L},\sqsubseteq)$ be a partially ordered set. If for
  each pair of elements $\ell,m\in\mathbb{L}$ there exists a supremum
  $\ell\sqcup m$ and an infimum $\ell\sqcap m$, we call
  $(\mathbb L,\sqcup,\sqcap)$ a \emph{lattice}. A \emph{bounded lattice} has a
  top element $1$ and a bottom element $0$. A lattice is
  \emph{complete} if every subset of $\mathbb{L}$ has an infimum and a
  supremum. It is \emph{distributive} if 
  $(\ell
  \sqcup m)\sqcap n=(\ell\sqcap n)\sqcup (m\sqcap n)$ holds for all
  $\ell,m,n\in\mathbb{L}$.

  A bounded lattice $\mathbb L$ is a \emph{Heyting algebra} if for any
  $\ell,m\in\mathbb L$, there is a greatest element $\ell'$ such that
  $\ell\sqcap \ell'\sqsubseteq m$. The residuum and negation are defined
  as $\ell\rightarrow m=\bigsqcup\{\ell'\mid \ell\sqcap \ell'\sqsubseteq m\}$
  and $\neg \ell=\ell \rightarrow 0$. A \emph{Boolean
    algebra} $\mathbb L$ is a Heyting algebra satisfying
  $\neg \neg \ell = \ell$ for all $\ell\in\mathbb L$.


\end{defn}
\begin{example}
  Given a set of atomic propositions $N$, consider $\mathbb{B}(N)$, the set of all Boolean
  expressions over $N$, i.e., the set of all formulae of propositional
  logic. We
  equate every subset $C\subseteq N$ with the evaluation that assigns
  $\mathit{true}$ to all $f\in C$ and $\mathit{false}$ to all
  $f\in N\backslash C$. For $b\in\mathbb{B}(N)$, we write $C\models b$
  whenever $C$ satisfies $b$. Furthermore we define
  $\llbracket b \rrbracket = \{C\subseteq N \mid C\models b\} \in
  \mathcal{P}(\mathcal{P}(N))$.
  Two Boolean expressions $b_1,b_2$ are called equivalent whenever
  $\llbracket b_1 \rrbracket = \llbracket b_2 \rrbracket$. Furthermore
  $b_1$ implies $b_2$ ($b_1\models b_2$), whenever
  $\llbracket b_1\rrbracket \subseteq \llbracket b_2\rrbracket$.

  The set $\mathbb{B}(N)$, quotiented by equivalence, is a Boolean
  algebra, isomorphic to $\mathcal{P}(\mathcal{P}(N))$, where
  $\llbracket b_1 \rrbracket \sqcup \llbracket b_2 \rrbracket =
  \llbracket b_1 \rrbracket \cup \llbracket b_2 \rrbracket =
  \llbracket b_1 \lor b_2 \rrbracket$,
  analogously for $\sqcap,\cap,\land$,
  $\lnot \llbracket b \rrbracket = \mathcal{P}(N)\backslash \llbracket
  b \rrbracket = \llbracket \lnot b \rrbracket$,
  and
  $\llbracket b_1 \rrbracket \to \llbracket b_2 \rrbracket =
  \mathcal{P}(N)\backslash \llbracket b_1 \rrbracket \cup \llbracket b_2 \rrbracket
  = \llbracket \lnot b_1 \lor b_2 \rrbracket$.
\end{example}
%
%

Distributive lattices and Boolean algebras give rise to an interesting
duality result, which was first stated for finite lattices by Birkhoff and
extended to the infinite case by Priestley \cite{dp:lattices-order}.
In the sequel we will focus on finite distributive lattices (which are Heyting algebras). 
We first need the following concepts.
%
\begin{defn}
  Let $\mathbb{L}$ be a lattice. An element $n\in\mathbb
  L\setminus\{0\}$ is said to be \emph{(join-)irreducible} if whenever
  $n=\ell\sqcup m$ for elements $\ell,m\in\mathbb L$, it always holds
  that $n=\ell$ or $n=m$. We write $\mathcal J(\mathbb L)$ for the set
  of all irreducible elements of $\mathbb L$.

  Let $(S,\le)$ be a partially ordered set. A subset
  $S'\subseteq S$ is \emph{downward-closed}, whenever $s'\in S'$ and $s\le s'$
  implies $s\in S'$. We write $\mathcal{O}(S)$ for the set of all
  downward-closed subsets of $S$ and $\history s = \{s' \mid s' \leq s\}$ for the downward-closure of $s\in S$.
\end{defn}
\begin{example}
  For our example of a Boolean algebra $\mathbb{B}(N)$, quotiented by
  equivalence, the irreducibles are the complete conjunctions of
  literals, or, alternatively, all sets $C\subseteq N$.
\end{example}

We can now state the Birkhoff's representation theorem for finite
distributive lattices~\cite{dp:lattices-order}.

\begin{theorem}
  \label{th:birkhoff}
  If $\mathbb L$ is a finite distributive lattice, then
  $(\mathbb L,\sqcup,\sqcap)\cong(\mathcal O(\mathcal J(\mathbb
  L)),\cup,\cap)$
  via the isomorphism
  $\eta:\mathbb L\rightarrow\mathcal O(\mathcal J(\mathbb L))$,
  defined as
  $\eta(\ell)=\{\ell'\in\mathcal J(\mathbb L)\mid \ell'\sqsubseteq \ell\}$.
  Furthermore, given a finite partially ordered set $(S,\leq)$, the
  downward-closed subsets of $S$, $(\mathcal O(S),\cup,\cap)$ form a
  distributive lattice, with inclusion ($\subseteq$) as the partial
  order. The irreducibles of this lattice are all downward-closed sets
  of the form $\history{s}$ for $s\in S$.
\end{theorem}

\begin{example}\label{ex:Lattice}
  Consider the lattice $\mathbb L=\{0,a,b,c,d,e,f,1\}$ with the order
  depicted in Figure~\ref{fig:ex:mot-Birkhoff}.
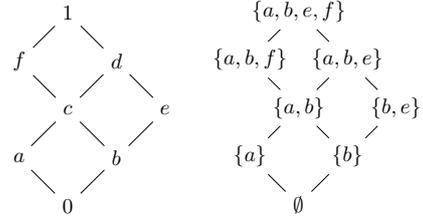
\begin{figure}
\centering
\scalebox{0.8}{
\begin{tikzpicture}[on grid, node distance=.8cm]
      \node (top) {$1$} ;
      \node[below=of top] (anker) {} ;
      \node[right= of anker] (ac) {$d$} ;
      \node[left= of anker] (d) {$f$} ;
      \node[below= of anker] (ab) {$c$} ;
      \node[below=of d] (anker2) {} ;
      \node[below= of anker2] (a) {$a$} ;
      \node[right= of a] (anker3) {} ;
      \node[right= of anker3] (b) {$b$} ;
      \node[right= of ab] (anker4) {} ;
      \node[right= of anker4] (c) {$e$} ;
      \node[below=of a] (anker5) {} ;
      \node[right= of anker5] (bot) {$0$} ;

      \begin{scope}[-]
        \draw (top) edge (d) ;
        \draw (top) edge  (ac) ;
        \draw (d) edge  (ab) ;
        \draw (ac) edge (c) ;
        \draw (ac) edge (ab) ;
        \draw (ab) edge (a) ;
        \draw (ab) edge (b) ;
        \draw (c) edge  (b) ;
        \draw (a) edge (bot) ;
        \draw (b) edge (bot) ;
      \end{scope}
    \end{tikzpicture}
    \quad
    \begin{tikzpicture}[on grid, node distance=.8cm]
      \node (top) {$\{a,b,e,f\}$} ;
      \node[below=of top] (anker) {} ;
      \node[right= of anker] (ac) {$\{a,b,e\}$} ;
      \node[left= of anker] (d) {$\{a,b,f\}$} ;
      \node[below= of anker] (ab) {$\{a,b\}$} ;
      \node[below=of d] (anker2) {} ;
      \node[below= of anker2] (a) {$\{a\}$} ;
      \node[right= of a] (anker3) {} ;
      \node[right= of anker3] (b) {$\{b\}$} ;
      \node[right= of ab] (anker4) {} ;
      \node[right= of anker4] (c) {$\{b,e\}$} ;
      \node[below=of a] (anker5) {} ;
      \node[right= of anker5] (bot) {$\emptyset$} ;

      \begin{scope}[-]
        \draw (top) edge (d) ;
        \draw (top) edge  (ac) ;
        \draw (d) edge  (ab) ;
        \draw (ac) edge (c) ;
        \draw (ac) edge (ab) ;
        \draw (ab) edge (a) ;
        \draw (ab) edge (b) ;
        \draw (c) edge  (b) ;
        \draw (a) edge (bot) ;
        \draw (b) edge (bot) ;
      \end{scope}
    \end{tikzpicture}}
    \caption{An example motivating Birkhoff's representation theorem.}\label{fig:ex:mot-Birkhoff}
    \vspace{-0.6cm}
\end{figure}
  \noindent
  The irreducible elements are $a,b,e,f$, i.e.\ exactly those elements
  that have a unique direct predecessor. On the right we depict the
  dual representation of the lattice in terms of downward-closed sets
  of irreducibles, ordered by inclusion.
This example suggests an embedding of a distributive lattice
$\mathbb{L}$ into a Boolean algebra, obtained by taking the powerset of
irreducibles.

\end{example}

\begin{proposition}[\rm Embedding]
  \label{prop:emb-lat-ba}
  A finite distributive lattice $\mathbb L$ embeds into the Boolean algebra $\mathbb B = \mathcal{P}(\mathcal J(\mathbb L))$ via the
  mapping $\eta:\mathbb L\rightarrow\mathbb B$ given by
  $\eta(\ell)=\{\ell'\in\mathcal J(\mathbb L)\mid \ell'\sqsubseteq \ell\}$.
\end{proposition}

We will simply assume that $\mathbb{L}\subseteq \mathbb{B}$.  Since an
embedding is a lattice homomorphism, supremum and infimum coincide in
$\mathbb L$ and $\mathbb B$ and we write $\sqcup,\sqcap$ for
both versions.  Negation and residuum may however differ and we
distinguish them via a subscript, writing
$\neg_{\mathbb L}, \neg_{\mathbb B}$ and
$\to_{\mathbb L},\to_{\mathbb B}$.  Given such an embedding, we can
approximate elements of a Boolean algebra in the embedded lattice.

\begin{defn}
  \label{def:approx}
  Let a complete distributive lattice $\mathbb L$ that embeds into
  a Boolean algebra $\mathbb B$ be given. Then, the
  \emph{approximation} of $\ell\in\mathbb B$ is given by: $\app{\ell}_\mathbb
  L=\bigsqcup\{\ell'\in\mathbb L\mid \ell'\sqsubseteq \ell\}.$
\end{defn}

If the lattice is clear from the context, we will in the sequel
drop the subscript $\mathbb{L}$ and simply write $\app{\ell}$. For
instance, in the previous example, the set of irreducibles
$\{a,e,f\}$, which is not downward-closed, is approximated by
$\app{\{a,e,f\}} = \{a\}$.



\begin{lemma}
  \label{lem:approximation}
  Let $\mathbb L$ be a complete distributive lattice that embeds into
  a Boolean algebra~$\mathbb B$.  For $\ell$, $m\in\mathbb{B}$, we
  have $\app{\ell\sqcap m}=\app{\ell}\sqcap\app{m}$ and furthermore that $\ell\sqsubseteq m$ implies
  $\app{\ell}\sqsubseteq \app{m}$. If
  $\ell,m\in\mathbb L$, then
  $\app{\ell\sqcup\neg m}=m\rightarrow_\mathbb L \ell$.
\end{lemma}

Note that in general it does not hold that $\app{\ell\sqcup
m}=\app{\ell}\sqcup\app{m}$ and $\app{\ell\sqcup\neg m}=\app{m}\rightarrow_\mathbb L
\app{\ell}$ for arbitrary $\ell,m\in\mathbb{B}$. To witness why these equations
fail to hold, take $\ell=\{a,e\}$ and $m=\{b,f\}$ in the
previous example as counterexample.



\section{Conditional Transition Systems}
\label{sec:cts}

In this section we introduce conditional transition systems together with a notion of
behavioural equivalence based on bisimulation. In \cite{ABHKMS12}, such transition systems were already investigated in a coalgebraic setting, where the set of conditions was trivially ordered. In the sequel, we will always use CTS for the variant with upgrades defined as follows:

\begin{defn}
  \label{CTS} A \emph{conditional transition system} (CTS) over an alphabet $A$ and a finite ordered set of conditions $(\Phi,\leq)$ is a triple $(X,A,f)$, where $X$ is a set of states and
  $f: X \times A \rightarrow (\Phi\rightarrow \mathcal{P}(X))$ is a function mapping
  every ordered pair in $X \times A$ to a monotone function of type
  $(\Phi,\leq) \rightarrow (\mathcal{P}(X),\supseteq)$. As usual, we write
  $x\xrightarrow{a,\phi} y$ whenever $y\in f(x,a)(\phi)$.
\end{defn}
%
Intuitively, a CTS evolves as follows. Before the system starts
acting, it is assumed that a condition $\phi\in \Phi$ is chosen
arbitrarily which may represent a selection of a valid product of the
system.
Now all the transitions that have a condition greater than or equal to
$\phi$ are activated, while the remaining transitions are inactive.
Henceforth, the system behaves like a standard transition system;
until at any point in the computation, the condition is changed to a
smaller one (say, $\phi'$) signifying a selection of a valid, upgraded
product.  This, in turn, has a propelling effect in the sense that now
(de)activation of transitions depends on the new condition $\phi'$,
rather than on the old condition $\phi$.  Note that due to the
monotonicity restriction we have that $x\xrightarrow{a,\phi} y$ and
$\phi'\le \phi$ imply $x\xrightarrow{a,\phi'} y$. That is, active
transitions remain active during an upgrade, but new transitions may
become active. In
\short{\cite{bkks:cts-upgrades-arxiv}}\full{Appendix~\ref{sec:deactivating-transitions}},
we weaken this requirement by discussing a mechanism for deactivating
transitions via priorities on the alphabet.
\begin{figure}
  \centering
  \scalebox{0.6}{
  \begin{tikzpicture}[state/.style={rounded rectangle,draw}]
    \node (i) {};
    \node[state] (r) at ($(i.center)+(1,0)$) {ready};
    \node[state] (rec) at ($(r.center)+(4,0)$) {received};
    \node[state] (rsafe) at ($(rec.center)+(1.5,2)$) {safe};
    \node[state] (runsafe) at ($(rec.center)+(1.5,-2)$) {unsafe};
    \path[->]
     (i) edge (r)
     (r) edge node[above] {receive,$\mathbf b$} (rec)
     (rec) edge node[below right] {check,$\mathbf b$} (rsafe)
     (rec) edge node[above right] {check,$\mathbf b$} (runsafe)
     (rsafe) edge node [above] {u,$\mathbf b$} (r)
     (runsafe) edge node [above right] {u,$\mathbf b$} (r)
     (runsafe) edge[bend left] node [below left] {e,$\mathbf a$} (r);
  \end{tikzpicture}
  }
  \caption{Adaptive routing protocol with the alphabet
    $A=\{\text{receive},\text{check},\text{u},\text{e}\}$.}\label{fig:protocol}
  \vspace{-0.6cm}
\end{figure}
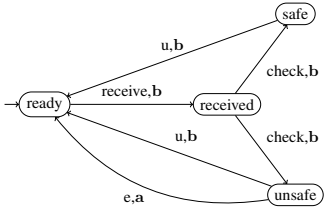

\begin{example}
  \label{ex:cts-bisim}
  Consider an example (simplified from
  \cite{Cordy2013:adaptivefts}) of an adaptive routing protocol
  modelled as a CTS in Figure~\ref{fig:protocol}.  The system has two
  products: the \emph{basic} system, denoted $\mathbf{b}$, with no
  encryption feature and the \emph{advanced} system, denoted
  $\mathbf{a}$, with an encryption feature. The ordering on the products
  is $\mathbf{a} < \mathbf{b}$.  Transitions that are
  present due to monotonicity are omitted.

  Initially, the system is in state 'ready' and is waiting to receive a
  message. 
  Once a message is received there is a check whether the system's
  environment is safe or unsafe, leading to non-deterministic
  branching.  If the encryption feature is present, then the system
  can send an encrypted message (e) from the unsafe state only;
  otherwise, the system sends an unencrypted message (u) regardless of
  the state being 'safe' or 'unsafe'. Note that such a behaviour description can easily be
  encoded by a transition function.  E.g.,
  $f(\text{received},\text{check})(\mathbf
  b)=\{\text{safe},\text{unsafe}\}$
  and $f(\text{received},a)(x)=\emptyset$ (for
  $x\in\{\mathbf a,\mathbf b\}$ and $a\in A\setminus\{\text{check}\}$)
  specifies the transitions that can be fired from the received state
  to the (un)safe states.
\end{example}

Next, we turn our attention towards (strong) bisimulation relations
for CTSs which consider the ordering of conditions in their transfer
properties. 
\begin{defn}\label{def:cts-bisim}
  Let $(X,A,f)$, $(Y,A,g)$ be two CTSs over the same set of conditions
  $(\Phi,\leq)$. For a condition $\phi\in\Phi$, we define
  $f_\phi(x,a)=f(x,a)(\phi)$ to denote the traditional
  \emph{($A$-)labelled transition system} induced by a CTS
  $(X,A,f)$. 
Two states $x\in X,y\in Y$ are \emph{conditionally bisimilar} under a
condition $\phi\in \Phi$, denoted $x \sim_\phi y$, if there is a
family of relations $R_{\phi'}\subseteq X\times Y$ (for every $\phi'\leq \phi$) such that
\begin{enumerate}[label=(\roman*)]
\item each relation $R_{\phi'}$ is a traditional bisimulation
  relation 
    between $f_{\phi'}$ and $g_{\phi'}$,
\item whenever $\phi'\leq \phi''$, we have $R_{\phi'}\supseteq R_{\phi''}$, and
\item $R_\phi$ relates $x$ and $y$, i.e., $(x,y)\in R_\phi$.
\end{enumerate}
\end{defn}

\begin{example}
  Consider the CTS
  illustrated in Figure~\ref{fig:protocol} where the condition
  $\mathbf b$ of the transition
  `$\text{received} \xrightarrow{\text{check}, \mathbf b} \text{unsafe}$'
  is replaced by $\mathbf a$. Let $\text{ready}_1$ and $\text{ready}_2$
  denote the initial states of the system before and after the above
  modification, respectively. Then, we find
  $\text{ready}_1 \sim_{\mathbf a} \text{ready}_2$; however,
  $\text{ready}_1 \not\sim_{\mathbf b} \text{ready}_2$. To see why the
  latter fails to hold, let $R_{\mathbf b}$ be the bisimulation relation
  in the traditional sense between the states
  $\text{ready}_1,\text{ready}_2$ under condition $\mathbf b$. Then,
  one finds that the states $\text{unsafe}_1,\text{safe}_2$ are
  bisimilar in the traditional sense, i.e.,
  $(\text{unsafe}_1,\text{safe}_2)\in R_{\mathbf b}$. However, the two
  states cannot be related by any traditional bisimulation relation
  under condition $\mathbf a$; thus, violating Condition 2 of
  Definition~\ref{def:cts-bisim}.


  Indeed, the two systems behave differently. In the first, it is
  possible to perform actions $\text{receive}$, $\text{check}$ (arrive
  in state $\text{unsafe}$), do an upgrade, and send an encrypted
  message ($\text{e}$), which is not feasible in the second system
  because the $\text{check}$ transition forces the system to be in the
  safe state before doing the upgrade. However, without upgrades, the
  above systems would be bisimilar for both products.
\end{example}
%
%
We end this section by adapting the classical bisimulation game to
conditional transition systems; thus, incorporating our intuitive explanation of upgrades with the notion of bisimilarity.

\begin{defn}[\rm Bisimulation Game]
  Given two CTSs $(X,A,f)$ and $(Y,A,g)$ over a poset $(\Phi, \leq)$,
  a state $x\in X$, a state $y\in Y$, and a condition $\phi\in\Phi$,
  the bisimulation game is a round-based two-player game that uses
  both CTSs as game boards.  Let $(x,y,\phi)$ be a game instance
  indicating that $x,y$ are marked and the current condition is
  $\phi$. The game progresses to the next game instance as follows:
  \begin{itemize}
  \item Player~1 is the first one to move. Player 1 can decide to make
    an upgrade, i.e., replace the condition $\phi$ by a smaller one (say $\phi'\leq\phi$, for some $\phi'\in\Phi$).
  \item Player~1 can  choose the marked state $x\in X$ (or
    $y\in Y$) and performs a transition $x\xrightarrow{a,\phi'}x'$
    ($y\xrightarrow{a,\phi'}y'$).
  \item Player~2 then has to simulate the last step, i.e., if Player~1 made a
    step $x\xrightarrow{a,\phi'}x'$, Player~2 is required to make step
    $y\xrightarrow{a,\phi'}y'$ and vice-versa.
  \item In turn, the new game instance is $(x',y',\phi')$.
  \end{itemize}
  Player~1 wins if Player~2 cannot simulate the last step performed by Player~1. Player~2 wins if the game never terminates or Player~1   cannot make another step.
\end{defn}

So bisimulation is characterised as follows: Player~2 has a winning
strategy for a game instance $(x,y,\phi)$ if and only if
$x\sim_\phi y$. The proof and the computation of the winning
strategies for both players are given in
\short{\cite{bkks:cts-upgrades-arxiv}}\full{Appendix~\ref{sec:proofs-strategies}}.

\section{Lattice Transition Systems}
\label{sec:lats}

Recall from Section~\ref{sec:preliminaries} that there is a duality
between partial orders and distributive lattices. In fact, as we will
show below, this result can be lifted to the level of transition
systems as follows: a conditional transition system over a poset is
equivalent to a transition system whose transitions are labelled by
the downward-closed subsets of the poset.

\begin{defn}
  \label{def:LatticeCTS}
  A \emph{lattice transition system} (LaTS) over a finite distributive lattice $\mathbb L$ and an alphabet $A$ is a triple
  $(X,A,\alpha)$ with a set of states $X$ and a transition
  function $\alpha:X \times A \times X \rightarrow \mathbb L$. A LaTS $(X,A,\alpha)$ is \emph{finite} if the sets $X,A$ are finite.
\end{defn}

Note that superficially, lattice transition systems resemble weighted
automata \cite{hwaDKV}. However, while in weighted automata the
lattice annotations are seen as weights that are accumulated, in CTSs
they play the role of guards that control which transitions can be
taken. Furthermore, the notions of behavioural equivalence are quite
different.

Given a CTS $(X,A,f)$ over $(\Phi,\le)$, we can easily construct a
LaTS over $\mathcal{O}(\Phi)$ by defining
$\alpha(x,a,x') = \{\phi\in\Phi \mid x'\in f(x,a)(\phi)\}$ for
$x,x'\in X$, $a\in A$. Due to monotonicity, $\alpha(x,a,x')$ is always
downward-closed. Similarly, a LaTS can be converted into a CTS by
using the Birkhoff duality and by taking the irreducibles as
conditions. 

\begin{theorem}
  The set of all CTSs over a set of conditions $\Phi$ is isomorphic to the set of all LaTSs over the lattice whose elements are
  the downward-closed subsets of $\Phi$.
\label{thm:CTSequiv}
\end{theorem}

So every LaTS over a finite distributive lattice gives rise to a CTS
in our sense (cf. Definition~\ref{CTS}) and since finite Boolean
algebras are finite distributive lattices, conditional transition
systems in the sense of \cite{ABHKMS12} are CTSs in our sense as
well. We chose the definition of a CTS using posets instead of the
dual view using lattices, because this view yields a natural
description which models transitions in terms of conditions (product versions), though when computing with CTSs we often choose the
lattice view. By adopting this view, conditional bisimulations can be
computed symbolically and hence more efficiently
(cf. Section~\ref{sec:bdd-based-repr}).

\begin{defn}
  \label{def:lattice-bisimulation}
  Let $(X,A,\alpha)$ and $(Y,A,\beta)$ be any two LaTSs over a lattice
  $\mathbb L$. A conditional relation $R$, i.e., a function of type
  $R: X\times Y \rightarrow \mathbb L$ is a \emph{lattice
    bisimulation} for $\alpha,\beta$ if and only if the following
  transfer properties are satisfied.
  \begin{enumerate}[label=(\roman*)]
  \item For all $x,x'\in X$, $y\in Y$, $a\in A$,
    $\ell\in \irreducibles$ whenever $x \xrightarrow{a,\ell} x'$ and
    $\ell \sqsubseteq R(x,y)$, there exists $y'\in Y$ such that
    $y\xrightarrow{a,\ell} y'$ and $\ell \sqsubseteq R(x',y')$.
    \item Symmetric to (i) with the roles of $x$ and $y$ interchanged.
  \end{enumerate}
  In the above, we write $x \xrightarrow {a,\ell} x'$, whenever
  $\ell \sqsubseteq \alpha(x,a,x')$.
\end{defn}

For $\phi\in\Phi$, a transition
$x \xrightarrow {a,\phi} x'$ exists in the CTS if and only if there is a transition
$x \xrightarrow {a,\history \phi}  x'$ in the corresponding LaTS.
Hence they are denoted by the same symbol.


\begin{theorem}
  \label{thm:bisim-correspondence}
  Let $(X,A,f)$ and $(Y,A,g)$ be any two CTSs over $\Phi$. Two states
  $x\in X,y\in Y$ are conditionally bisimilar under a condition $\phi$
  if and only if there is a lattice bisimulation $R$
  between the corresponding LaTSs such that $\phi \in R(x,y)$.
\end{theorem}

Incidentally, the order in $\mathbb L$ gives rise to a natural
order on lattice bisimulations. For any two lattice
bisimulations $R_1, R_2: X\times Y\rightarrow\mathbb L$, we write $R_1\sqsubseteq R_2$ if and only if
$R_1(x,y)\sqsubseteq R_2(x,y)$ for all $x\in X,y\in Y$. As a result,
taking the element-wise supremum of a family of lattice bisimulations
is again a lattice bisimulation. Therefore, the greatest lattice
bisimulation for a LaTS always exists, just like in the traditional
case.

\begin{lemma}
  \label{lem:unionisbisim}
  Let $R_i\in X\times Y\rightarrow\mathbb L, i\in I$ be lattice
  bisimulations for a pair of LaTSs $(X,A, \alpha)$ and $(Y,A,\beta)$.
  Then $\bigsqcup\{R_i\mid i\in I\}$ is a lattice bisimulation.
\end{lemma}



\section{Computation of Lattice Bisimulation}

\label{sec:matrix-mult}
The goal of this section is to present an algorithm that computes the
greatest lattice bisimulation between a given pair of LaTSs. In
particular, we first characterise lattice bisimulation as a
post-fixpoint of an operator $F$ on the set of all conditional
relations. Then, we show that this operator $F$ is monotone with
respect to the ordering relation $\sqsubseteq$; thereby, ensuring that
the greatest bisimulation always exists by applying the well-known
Knaster-Tarski fixpoint theorem. Moreover, on finite lattices and
finite sets of states, the usual fixpoint iteration starting with the
trivial conditional relation (i.e., the constant $1$-matrix over $\mathbb L$) can be used to compute the greatest lattice bisimulation. Lastly, we give a translation of $F$ in terms of matrices using a form of matrix multiplication found in the literature of residuated lattices \cite{Belohlavek:2012:matrix-multi} and database design \cite{Kohout:1985:matrix-mult}.

\subsection{A Fixpoint Approach}

Throughout this section, we let $\alpha: X\times A\times X\rightarrow\mathbb L$, $\beta: Y\times A\times Y\rightarrow\mathbb L$ denote any two LaTSs, $\mathbb L$ denote a finite distributive lattice, and $\mathbb B$ denote the Boolean algebra that this lattice embeds into.
\begin{defn}
  \label{def:F-operator}
  Recall the residuum operator $\to$ on a lattice and define three operators $F_1,F_2,F:(X\times Y\rightarrow\mathbb L)\rightarrow(X\times
  Y\rightarrow\mathbb L)$ in the following way (for $R\in X\times Y\rightarrow\mathbb L$, $x\in X$, $y\in Y$):
{
\allowdisplaybreaks
\begin{align*}
  &F_1(R)(x,y) = \\
  &\bigsqcap_{a\in A,x'\in X}\bigg(\alpha(x,a,x') \rightarrow
                \big(\bigsqcup_{y'\in Y}(\beta(y,a,y')\sqcap
                R(x',y'))\big) \bigg),\\
\full{
  &F_2(R)(x,y) = \\
  &\bigsqcap_{a\in A,y'\in Y}\bigg(\beta(y,a,y') \rightarrow
  \big(\bigsqcup_{x'\in X}(\alpha(x,a,x')\sqcap
  R(x',y'))\big)\bigg),\\
}
  &F(R)(x,y) = F_1(R)(x,y) \sqcap F_2(R)(x,y).
\end{align*}
}\short{The operator $F_2$ is defined analogously to $F_1$ where the
  roles of $x,y$ as well as $x',y'$ and $\alpha,\beta$ are
  interchanged.}
\end{defn}

Note that the above definition is provided for a distributive lattice,
viewing it in classical two-valued Boolean algebra results in the well-known transfer properties of a bisimulation.

\begin{theorem}
  \label{thm:fixpoint-lattice-bisim}
  A conditional relation $R$ is a lattice bisimulation if and only if $R\sqsubseteq F(R)$.
\end{theorem}

Next, it is easy to see that $F$ is a monotone operator with respect
to the ordering $\sqsubseteq$ on $\mathbb L$ since the infimum and
supremum are both monotonic, and moreover, the residuum operation is
monotonic in the second component.
As a result, we can use the following fixpoint iteration to compute
the greatest bisimulation while working with finite lattices and
finite sets of states.

\begin{myalgorithm}\label{algo:partref}
  Let $(X,A,\alpha)$ and $(Y,A,\beta)$ be two finite LaTSs. Fix $R_0$ as $R_0(x,y)=1$ for all
  $x\in X,y\in Y$. Then, compute $R_{i+1}=F(R_i)$ for all
  $i\in\mathbb N_0$ until $R_{i}\sqsubseteq R_{i+1}$. Lastly, return
  $R_i$ as the greatest bisimulation.
\end{myalgorithm}
Suppose $\alpha=\beta$, it is not hard to see that the fixpoint
iteration must stabilise after at most $|X|$ steps, since each $R_i$
induces equivalence relations for all conditions $\phi$ and
refinements regarding $\phi$ are immediately propagated to every
$\phi'\ge \phi$. An equivalence relation can be refined at most $|X|$ times, limiting the number of iterations.

\subsection{Lattice Bisimilarity is Finer than Boolean Bisimilarity}

We now show the close relation of the notions of bisimilarity for a
LaTS defined over a finite distributive lattice $\mathbb L$ and a
Boolean algebra $\mathbb B$.
%
As usual, let $(X,A,\alpha)$ and $(Y,A,\beta)$ be any two LaTSs
together with the restriction that the lattice $\mathbb L$ embeds into
the Boolean algebra $\mathbb B$. Moreover, let $F_{\mathbb L}$ and
$F_{\mathbb B}$ be the monotonic operators as defined in
Definition~\ref{def:F-operator} over the lattice $\mathbb L$ and the
Boolean algebra $\mathbb B$, respectively. We say that $R$ is an
$\mathbb{L}$-bisimulation (resp. $\mathbb B$-bisimulation) whenever $R\sqsubseteq F_{\mathbb L}(R)$ (resp. $R \sqsubseteq F_{\mathbb B}(R)$).


\begin{proposition}\label{prop:lat-bisim&bool-bisim}
  \mbox{}
  \begin{enumerate}[label=(\roman*)]
  \item If $R:X\times Y\rightarrow\mathbb L$, then
    $\app{F_{\mathbb B}(R)} = F_\mathbb L(R)$.
  \item Every $\mathbb L$-bisimulation is also a
    $\mathbb B$-bisimulation.
  \item A $\mathbb B$-bisimulation $R:X\times Y\rightarrow\mathbb B$
    is an $\mathbb L$-bisimulation whenever all the entries of $R$ are
    in $\mathbb L$.
  \end{enumerate}
\end{proposition}

However, even though the two notions of bisimilarity are closely
related, they are not identical, i.e., it is not true that whenever a state $x$ is bisimilar to a state $y$ in $\mathbb B$ that it is also
bisimilar in $\mathbb L$ (see Example~\ref{ex:cts-bisim} where we
encounter a $\mathbb{B}$-bisimulation, which is not an $\mathbb{L}$-bisimulation).

\subsection{Matrix Multiplication}

An alternative way to represent a LaTS $(X,A,\alpha)$ is to view the transition function $\alpha$ as a family of matrices
$\alpha_a: X\times X \rightarrow \mathbb L$ (one for each action
$a\in A$) with 
$\alpha_a(x,x')=\alpha(x,a,x')$, for every $x,x'\in X$.
We use standard matrix multiplication (where $\sqcup$ is used for
addition and $\sqcap$ for multiplication), as well as a special form
of matrix multiplication
\cite{Belohlavek:2012:matrix-multi,Kohout:1985:matrix-mult}.

\begin{defn}[\rm $\otimes$-multiplication]
  Given an $X\times Y$-matrix $U\colon X\times Y\to\mathbb{L}$ and a
  $Y\times Z$-matrix $V\colon Y\times Z\to \mathbb{L}$, we define
  the $\otimes$-multiplication of $U$ and $V$ as follows:
  \full{\[ U\otimes V\colon X\times Z \to \mathbb{L} \]}
  \[(U\otimes V)(x,z)=\bigsqcap_{y\in Y} \big(U(x,y)\rightarrow_{\mathbb L} V(y,z)\big)
  \enspace.\]
\end{defn}
%
\begin{theorem}\label{thm:bisim-imp}
  Let $R:X\times Y \rightarrow L$ be a conditional relation between a   pair of LaTSs $(X,A,\alpha)$ and $(Y,A,\beta)$. Then,
  $F(R)= \bigsqcap_{a\in
    A}((\alpha_a\otimes(R\cdot{\beta_a}^T))\sqcap(\beta_a\otimes(\alpha_a\cdot
  R)^T)^T)$,
  where $A^T$ denotes the transpose of a matrix $A$.
\end{theorem}

We end this section by making an observation on LaTSs over
a Boolean algebra. In a Boolean algebra, it is well-known that the
residuum operator can be replaced by the negation and join
operators. Thus, in this case, using only the standard matrix multiplication and (componentwise) negation we get $U\otimes V = \lnot (U\cdot (\lnot V))$ which further simplifies $F(R)$ as: 
\[F(R)=\bigsqcap_{a\in A}\bigl(\neg(\alpha_a \cdot \neg(R \cdot \beta_a^T)) \sqcap \neg(\neg(\alpha_a \cdot R) \cdot \beta_a^T)\bigr) \enspace.\]
This reduction is especially relevant to software product lines with no upgrade features.

\section{Application and Implementation}
\label{sec:spl}

\subsection{Featured Transition Systems}

A Software Product Line (SPL) is commonly described as ``a set of
software-intensive systems that share a common, managed set of
features satisfying the specific needs of a particular market segment
or mission and that are developed from a common set of core assets
[artifacts] in a prescribed way'' \cite{2001:SPL:501065}. The idea of
designing a set of software systems that share common functionalities
in a collective way is becoming prominent in the field of software
engineering (cf. \cite{Metzger:2014:SPL:2593882.2593888}). In this
section we show that a featured transition system (FTS)
\cite{DBLP:conf/icse/CordyCPSHL12} -- a
well-known formal model that is expressive enough to specify an SPL --
is a special instance of a CTS.
\begin{defn}
  A \emph{featured transition system} (FTS) over a finite set of
  \emph{features} $N$ is a tuple $\mathcal F = (X,A, T, \gamma)$,
  where $X$ is a finite set of states, $A$ is a finite set of actions
  and $T\subseteq X\times A \times X$ is the set of transitions.
  Finally, $\gamma: T\rightarrow \mathbb{B}(N)$ assigns a Boolean
  expression over $N$ to each transition.
\end{defn}

FTSs are often accompanied by a so-called \emph{feature diagram} \cite{Cordy2013:adaptivefts,Classen:2010:MCL:1806799.1806850,Classen:2013:FTS},
a Boolean expression $d\in \mathbb{B}(N)$
that specifies admissible feature combinations.
Given a subset of features $C\subseteq N$ (called \emph{configuration}
or \emph{product}) such that $C\models d$ and an FTS
$\mathcal F=(X,A,T,\gamma)$, a state $x\in X$ can perform an
$a$-transition to a state $y\in X$ in the configuration $C$, whenever
$(x,a,y)\in T$ and $C\models \gamma(x,a,y)$.

It is easy to see that an FTS is a CTS, where the
conditions are subsets of $N$ satisfying $d$ with the discrete
order.
Moreover, an FTS is a special case of an LaTS due to Theorem~\ref{thm:CTSequiv}
and $\mathcal O(\llbracket d \rrbracket,=) = \mathcal P(\llbracket d \rrbracket)$. Given an FTS $\mathcal F = (X, A, T, \gamma)$ and a feature diagram $d$, then the corresponding LaTS is $(X, A, \alpha)$ with $\alpha(x,a,y)=\llbracket \gamma(x,a,y)\land d\rrbracket$, if $(x,a,y)\in T$; $\alpha(x,a,y)=\emptyset$, if $(x,a,y)\not\in T$.

Furthermore, we can extend the notion of FTSs by fixing a subset of
upgrade features $U\subseteq N$ that induces the following ordering on
configurations $C,C'\in \llbracket d\rrbracket$:
\begin{align*}
  C\le C' \iff \ & \forall f\in U(f\in C'\Rightarrow f\in C) \ \land \\
  & \forall f\in (N\backslash U)\, (f\in C' \iff f\in C).
\end{align*}

Intuitively, the configuration $C$ can be obtained from $C'$ by
``switching'' on one or several upgrade features $f\in U$. Notice that
it is this upgrade ordering on configurations which gives rise to the
partially ordered set of conditions in the definition of a CTS.
Hence, in the following we will consider the lattice
$\mathcal O(\llbracket d \rrbracket, \le)$ (i.e., the set of all
downward-closed subsets of $\llbracket d \rrbracket$).






\subsection{BDD-Based Representation}
\label{sec:bdd-based-repr}

In this section, we discuss our implementation of lattice bisimulation
using a special form of binary decision diagrams (BDDs) called
\emph{reduced and ordered binary decision diagrams} (ROBDDs). Our
implementation can handle adaptive SPLs that allow upgrade
features, using finite distributive lattices. Note that non-adaptive SPLs based
on Boolean algebras are a special case. BDD-based implementations of
FTSs without upgrades have already been mentioned in
\cite{DBLP:conf/icse/CordyCPSHL12,Classen:2011:symbolic}.

A \emph{binary decision diagram} (BDD) is a rooted, directed, and acyclic graph which serves as a representation of a Boolean function.
Every BDD has two distinguished terminal nodes $1$ and $0$,
representing the logical constants \emph{true} and \emph{false}.  The
inner nodes are labelled by the atomic propositions of a Boolean
expression $b\in\mathbb{B}(N)$ represented by the BDD, such that on
each path from the root to the terminal nodes, every variable of the
Boolean formula occurs at most once. Each inner node has exactly two
distinguished outgoing edges called \emph{high} and \emph{low}
representing the case that the atomic proposition of the inner node
has been set to \emph{true} or \emph{false}. Given a BDD for a Boolean
expression $b\in\mathbb{B}(N)$ and a configuration $C\subseteq N$
(representing an evaluation of the atomic propositions), we can check
whether $C\models b$ by following the path from the root node to a
terminal node. At a node labelled $f\in N$ we go to the
\textit{high}-successor if $f\in C$ and to the \emph{low}-successor if
$f\not\in C$. If we arrive at the terminal node labelled $1$ we have
established that $C\models b$, otherwise $C\not\models b$.

We use a special class of BDDs called ROBDDs (see \cite{And97} for more details) in which the order
of the variables occurring in the BDD is fixed and redundancy is
avoided.
If both the child nodes of a parent node are identical, the
parent node is dropped from the BDD and isomorphic parts of the BDD
are merged.
The advantage of ROBDDs is that two equivalent Boolean
formulae are represented by exactly the same ROBDD (if the order of
the variables is fixed).
Furthermore, there are polynomial-time
implementations for the basic operations -- negation, conjunction, and
disjunction.
These are however sensitive to the ordering of atomic
propositions and an exponential blowup cannot be ruled out, but often
it can be avoided.
\begin{wrapfigure}{R}{0.17\textwidth}
  \centering
  \vspace{-0.2cm}%
  \scalebox{0.7}{
\def\boundb{(-3.5,-5.2) rectangle (1.5,.5)}
\scalebox{.8}{
\begin{tikzpicture}[node distance=1.1 and 0.9, on grid, shorten >=1pt, >=stealth', semithick]

	\begin{scope}[circle, inner sep=3pt, minimum size=0pt]
		\draw node [draw] (b0) {\(f_0\)};
		\draw node [draw, below left=of b0] (b1a) {\(f_1\)};
		\draw node [draw, below right=of b0] (b1b) {\(f_1\)};
		\draw node [draw, below left=of b1a] (b2) {\(f_2\)};
		\draw node [draw, below left=of b2] (b3a) {\(f_3\)};
		\draw node [draw, below right=of b2] (b3b) {\(f_3\)};
	\end{scope}

	\begin{scope}[rectangle, inner sep=3pt, minimum size=0pt]
		\draw node [draw, below=of b3a] (1) {\(1\)};
		\draw node [draw, below=of b3b] (0) {\(0\)};
	\end{scope}

	\begin{scope}[->]
		\draw (b0) edge [dashed] (b1a);
		\draw (b0) edge (b1b);
		\draw (b1a) edge [dashed] (b2);
		\draw (b1a) edge [bend left] (0);
		\draw (b1b) edge [dashed, bend left] (0);
		\draw (b1b) edge (b2);
		\draw (b2) edge [dashed] (b3a);
		\draw (b2) edge (b3b);
		\draw (b3a) edge [dashed] (1);
		\draw (b3a) edge (0);
		\draw (b3b) edge [dashed] (0);
		\draw (b3b) edge (1);
	\end{scope}

\end{tikzpicture}}}
  \caption{BDD for $b$.
  }
  \label{fig:BDD}
  \vspace{-0.4cm}
\end{wrapfigure}
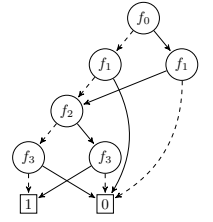

Consider a Boolean expression $b$ with
$\llbracket b\rrbracket = \{\emptyset, \{f_2,f_3\},$ $ \{f_0,f_1\},
\{f_0,f_1, f_2, f_3\}\}$
and the ordering on the atomic propositions as $f_0,f_1,f_2,f_3$.
Figure~\ref{fig:BDD} shows the corresponding ROBDD representation for $b$, where the inner
nodes, terminal nodes, and high (low) edges are depicted as circles,
rectangles, and solid (dashed) lines, respectively.

Formally, an ROBDD $b$ over a set of features $N$ is an
expression in one of the following forms: $ 0$, or $1$, or
$(f,b_1,b_0)$.  Here, $0,1$ denote the two terminal nodes and the
triple $(f,b_1,b_0)$ denotes an inner node with variable $f\in N$ and
$b_0,b_1$ as the \emph{low}- and \emph{high}-successors,
respectively. If $b=(f,b_1,b_0)$, we write $\mathit{root}(b) = f$,
$\mathit{high}(b) = b_1$, and $\mathit{low}(b) = b_0$.

Note that the elements of the Boolean algebra
$\mathcal{P}(\mathcal{P}(N))$ {\it correspond exactly} to ROBDDs over
$N$.  We now discuss how ROBDDs can be used to specify and
manipulate elements of the lattice
$\mathcal{O}(\llbracket d\rrbracket,\le)$.  In particular, computing
the infimum (conjunction) and the supremum (disjunction) in the
lattice $\mathcal{O}(\llbracket d\rrbracket,\le)$ is standard, since
this lattice can be embedded into $\mathcal{P}(\mathcal{P}(N))$ and
the infimum and supremum operations coincide in both structures.
Thus, it remains to characterize the lattice elements and the
residuum operation. 

We say that an ROBDD $b$ is {\em downward-closed} w.r.t.
$\le$ (or simply, downward-closed) if the set of configurations
$\llbracket b \rrbracket$ is downward-closed w.r.t. $\le$.
The following lemma characterises when an ROBDD $b$ is
downward-closed. It follows from the fact that
$F\in \mathcal{P}(\mathcal{P}(N))$ is downward-closed if and only if
for all $C\in F, f\in U$ we have $C\cup \{f\}\in F$.

\begin{lemma}
  \label{lem:highsmallerlow}
  An ROBDD  is downward-closed if and only if for each node labelled with a upgrade feature, the \emph{low}-successor implies the \emph{high}-successor.
\end{lemma}

Next, we compute the residuum in
$\mathcal{O}(\llbracket d\rrbracket,\le)$ by using the residuum
operation of the Boolean algebra $\mathcal{P}(\mathcal{P}(N))$.  For
this, we first describe how to approximate an element of the Boolean
algebra (represented as an ROBDD) in the lattice $\mathcal{O}(\mathcal{P}(N),\le)$.

\begin{algorithm}\caption{Approximation $\lfloor\!\!\lfloor b \rfloor\!\!\rfloor$ of an ROBDD $b$ in the lattice $\mathcal{O}(\mathcal{P}(N),\le)$}
  \label{alg:approxROBDD}
  \mbox{}

  \textbf{Input: } An ROBDD $b$ over a set of features $N$ and a set of upgrade features $U\subseteq N$.

  \textbf{Output: } An ROBDD $\approxL b$, which is the best approximation of  $b$ in the lattice. 

\begin{algorithmic}[1]
\Procedure{$\approxL b$}{}
\If {$b$ is a leaf} \Return {$b$} \ElsIf {$root(b) \in U$}  \Return \\
{\qquad\qquad $\textit{build}(\mathit{root}(b), \approxL{\mathit{high}(b)}, \approxL{\mathit{high}(b)}\wedge\approxL{\mathit{low}(b)})$}
\Else { \Return{$\mathit{build}(\mathit{root}(b), \approxL{\mathit{high}(b)}, \approxL{\mathit{low}(b)})$}}
\EndIf
\EndProcedure
\end{algorithmic}

\end{algorithm}

In the above algorithm, for
each non-terminal node that carries a label in $U$ (line $3$), we replace the
\emph{high}-successor with the conjunction of the \emph{low} and the
\emph{high}-successor using the procedure described above. Since
this might result in a BDD that is not reduced, we apply the
$\mathit{build}$ procedure appropriately, which
simply transforms a given ordered BDD into an ROBDD.
The result of the algorithm $\approxL b$ coincides with the approximation $\app{b}$ of the ROBDD $b$ seen as an element of the Boolean algebra $\mathcal{P}(\mathcal{P}(N))$ (Definition~\ref{def:approx}).
\begin{lemma}
  \label{lem:approximationROBDD}
  For an ROBDD $b$, $\approxL b$ is downward-closed.  Furthermore,
  $\approxL b\models b$ and there is no other downward-closed ROBDD
  $b'$ such that $\approxL b\models b' \models b$.  Hence
  $\approxL{b} = \app{b}$.
\end{lemma}
For each node in the BDD we compute at most one supremum, which is
quadratic. Hence the entire runtime of the approximation procedure is
at most cubic.
%
%
%
%
Finally, we discuss how to compute the residuum in
$\mathcal{O}(\llbracket d\rrbracket,\le)$.

\begin{proposition}
  \label{prop:residuuminOd}
  Let $b_1,b_2$ be two ROBDD which represent elements of
  $\mathcal{O}(\llbracket d\rrbracket,\le)$, i.e., $b_1,b_2$ are both
  downward-closed and $b_1\models d$, $b_2\models d$. (i)
  $\app{\lnot b_1\lor b_2\lor \lnot d}\land d$ is the residuum
  $b_1\to b_2$ in the lattice
  $\mathcal{O}(\llbracket d\rrbracket,\le)$.  (ii) If $d$ is
  downward-closed, then this simplifies to
  $b_1\to b_2 = \app{\lnot b_1\lor b_2}\land d$.

  Here, $\lnot$ is the negation in the Boolean algebra $\mathcal{P}(\mathcal{P}(N))$.
\end{proposition}

\subsection{Implementation and Runtime Results}
\label{sec:impl-runtime}


We have implemented an algorithm that computes the lattice
bisimulation relation based on the matrix multiplication (see
Theorem~\ref{thm:bisim-imp}) in a generic way.  Specifically, this
implementation is independent of how the irreducible elements are
encoded, ensuring that no implementation details of operations such as
matrix multiplication can interfere with the runtime results.  For our
experiments we instantiated it in two possible ways: with bit vectors
representing feature combinations and with ROBDDs as outlined above.
Our results show a significant advantage when we use BDDs to compute
lattice bisimilarity.
The implementation is written in C\# and uses the CUDD package by
Fabio Somenzi via the interface PAT.BDD \cite{NguyenSLDL12}.

\begin{figure}
  \centering
\scalebox{0.7}{
  \raisebox{100pt}{$\alpha:$}\hspace{-0.5cm}
  \begin{tikzpicture}[x={(1.2cm,-.7cm)},y={(0,2cm)},z={(1.8cm,.7cm)}]
    \node[state] (q1) {$0$} ;
    \node[right=of q1] (anker) {} ;
    \node[state,above= of anker] (q3) {$2$} ;
    \node[state,below= of anker] (q2) {$1$} ;

    \begin{scope}[->]
      \draw (q1) edge node [above right]{$b, \llbracket f \rrbracket$} (q2) ;
      \draw[loop left] (q1) edge node [above]{$b, \llbracket f \rrbracket$} (q1) ;
      \draw[bend right] (q1) edge node [right] {$b,\llbracket\mathit{true}\rrbracket$} (q3) ;
      \draw[bend right] (q3) edge node [left] {$c, \llbracket f \rrbracket$} (q1) ;
    \end{scope}
  \end{tikzpicture}
  \quad
  \raisebox{100pt}{$\beta:$}\hspace{-0.5cm}
  \begin{tikzpicture}[x={(1.2cm,-.7cm)},y={(0,2cm)},z={(1.8cm,.7cm)}]
    \node[state] (q1) {$0$} ;
    \node[right=of q1] (anker) {} ;
    \node[state,above= of anker] (q3) {$2$} ;
    \node[state,below= of anker] (q2) {$1$} ;

    \begin{scope}[->]
      \draw (q1) edge node [above right]{$b, \llbracket f \rrbracket$} (q2) ;
      \draw[loop left] (q1) edge node [above]{$b, \llbracket f \rrbracket$} (q1) ;
      \draw[bend right] (q1) edge node [right] {$b,\llbracket f \rrbracket$} (q3) ;
      \draw[bend right] (q3) edge node [left] {$c, \llbracket f \rrbracket$} (q1) ;
    \end{scope}
  \end{tikzpicture}}
		\caption{Components for $\alpha$ and $\beta$, where
                  $f$ is viewed as a Boolean expression indicating the
                  presence of feature $f$.}
\vspace{-0.6cm}
	\label{ex:ctscomp}
\end{figure}
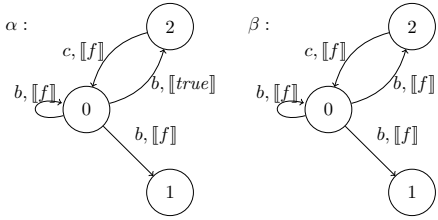
To show that the use of BDDs can potentially lead to an
exponential gain in speed when compared to the naive bit-vector
implementation, we executed the algorithm on a family of increasingly
larger LaTSs over an increasingly larger number of features, where all
features are upgrade features.
Let $F$ be a set of features.  Our example contains,
for each feature $f\in F$, one disconnected component in both LaTSs
that is depicted in Figure~\ref{ex:ctscomp}: the component for
$\alpha$ on the left, the component for $\beta$ is on the right.
The only difference between the two is in the guard of the transition from state $0$ to state $2$.

The quotient of the times taken without BDDs and with BDDs is growing
exponentially by a factor of about~$2$ for each additional feature
\short{(see the runtime results in
  \cite{bkks:cts-upgrades-arxiv})}\full{(see the table in
  Appendix~\ref{sec:runtime-results})}. Due to fluctuations, an exact
rate cannot be given. By the eighteenth iteration (i.e. 18~features
and copies of the basic component), the implementation using BDDs
needed 17~seconds, whereas the version without BDDs took more than
96~hours.  The nineteenth iteration exceeded the memory for the
implementation without BDDs, but terminated within 22~seconds with
BDDs.

\section{Conclusion, Related Work, and Future Work}
\label{sec:conclusion}

In this paper, we endowed CTSs with an order on conditions to model systems whose behaviour can be upgraded by replacing the current condition by a smaller one.
Corresponding verification techniques based on behavioural
equivalences can be important for SPLs where an upgrade to a more
advanced version of the same software should occur without unexpected
behaviour.
%
To this end, we proposed an algorithm, based on matrix multiplication,
that allows to compute the greatest bisimulation of two given CTSs.
Interestingly, the duality between lattices and downward-closed sets
of posets, as well as the embedding into a Boolean algebra proved to
be fruitful when developing it and proving its correctness.

There are two ways in which one can extend CTSs as a specification
language: first, in some cases it makes sense to specify that an
advanced version offers improved transitions with respect to a basic
version. For instance, in our running example, allowing the router to
send unencrypted messages in an unsafe environment is superfluous
because the advanced version always has the encryption feature. Such a
situation can be modelled in a CTS by adding a precedence relation
over the set of actions, leading to the deactivation of transitions,
which is worked out in
\short{\cite{bkks:cts-upgrades-arxiv}}\full{Appendix~\ref{sec:deactivating-transitions}}.
The second question is how to incorporate downgrades: one solution
could be to work with a pre-order on conditions, instead of an order.
This simply means that two conditions $\phi\neq\psi$ with
$\phi\le \psi$, $\psi\le\phi$ can be merged since they can be
exchanged arbitrarily.  Naturally, one could study more sophisticated
notions of upgrade and downgrade in the context of adaptivity.


As for the related work on adaptive SPLs, literature can be grouped
into either empirical or formal approaches; however, given the nature of our work, below we rather
concentrate only on the formal ones
\cite{Cordy2013:adaptivefts,Chrszon2016:profeat,Dubslaff:2014:PMC,terBeek:2015:statistical}.

Cordy et al. \cite{Cordy2013:adaptivefts} model an adaptive SPL using
an FTS which encodes not only a
product's transitions, but also how some of the features may change via the execution
of a transition.
In contrast, we encode adaptivity by requiring a partial order on the
products of an SPL and its effect on behaviour evolution by the
monotonicity requirement on the transition function. Moreover, instead
of studying the model checking problem as in
\cite{Cordy2013:adaptivefts}, our focus was on bisimilarity between
adaptive SPLs.

In \cite{Dubslaff:2014:PMC,Chrszon2016:profeat,Lochau2017},
alternative ways to model adaptive SPLs by using the synchronous
parallel composition on two separate computational models is
presented. Intuitively, one models the static aspect of an SPL, while
the other focuses on adaptivity by specifying the dynamic
(de)selection of features.  For instance, Dubslaff et
al. \cite{Dubslaff:2014:PMC} used two separate Markov decision
processes (MDP) to model an adaptive SPL. They modelled the core
behaviour in an MDP called \emph{feature module}; while dynamic
(de)activation of features is modelled separately in a MDP called
\emph{feature controller}. In retrospect, our work shows that for
monotonic upgrades it is possible to \emph{compactly} represent an
adaptive SPL over one computational model (CTSs in our case) rather
than a parallel composition of two.


In \cite{terBeek:2015:statistical}, a process calculus QFLan motivated
by concurrent constraint programming was developed.  Thanks to an
in-built notion of a store, various aspects of an adaptive SPL such as
(un)installing a feature and replacing a feature by another feature
can be modelled at run-time by operational
rules. Although QFLan has constructs to specify
quantitative constraints in the spirit of
\cite{Dubslaff:2014:PMC}, their aim is to obtain
statistical evidence by performing simulations.

Behavioural equivalences such as (bi)simulation relations have already been studied in the literature of traditional SPLs. In \cite{DBLP:conf/icse/CordyCPSHL12}, the authors proposed a
definition of simulation relation between any two FTSs (without upgrades) 
to combat the state explosion problem by establishing a simulation relation between a system and its refined version.
In contrast, the authors in \cite{Atlee:2015:MBI:2820126.2820133} used simulation relations to
measure the discrepancy in behaviour caused by feature interaction,
i.e., whether a feature that is correctly designed in isolation works correctly when combined with the other features or not.

(Bi)simulation relations on lattice Kripke structures were also
studied in \cite{doi:10.1142/S0129054110007192},
but in a very different context (in model-checking rather than in the
analysis of adaptive SPLs).  Disregarding the differences between
transition systems and Kripke structures (i.e., forgetting the role of
atomic propositions), the definition of bisimulation in
\cite{doi:10.1142/S0129054110007192} is quite similar to our
Definition~\ref{def:F-operator} (another similar formula occurs in
\cite{DBLP:conf/icse/CordyCPSHL12}).  However, in
\cite{doi:10.1142/S0129054110007192} the stronger
assumption 
of finite distributive de Morgan algebras is used, the results are
quite different and symbolic representations via BDDs are not taken
into account. Moreover, representing the lattice elements and computing residuum over them using the BDDs is novel in comparison with \cite{DBLP:conf/icse/CordyCPSHL12,doi:10.1142/S0129054110007192}.

Lastly, Fitting \cite{DBLP:conf/aiml/Fitting02} studied bisimulation
relations in the setting of unlabelled transition systems and gave an
elegant characterisation of bisimulation when transition systems and
the relations over states are viewed as matrices. By restricting
ourselves to LaTSs over Boolean algebras and fixing our alphabet to be
a singleton set, we can establish the following correspondence between
Fitting's formulation of bisimulation and lattice bisimulation (see
\short{\cite{bkks:cts-upgrades-arxiv}}\full{Appendix~\ref{sec:proofs-lattices}}
for the proof).

\begin{theorem}\label{thm:fitting}
  Let $(X,\alpha)$ be a LaTS over an atomic Boolean algebra
  $\mathbb B$. Then, a conditional relation
  $R:X\times X \rightarrow \mathbb B$ is a lattice bisimulation for
  $\alpha$ if and only
  if 
  $R\cdot\alpha \sqsubseteq \alpha \cdot R$ and
  $R^T \cdot \alpha \sqsubseteq \alpha \cdot R^T$.  Here we interpret
  $\alpha$ as a matrix of type $X \times X \rightarrow \mathbb L$ by
  dropping the occurrence of action labels.
\end{theorem}

In hindsight, we are treating general
distributive lattices that allow us to conveniently model and reason
about upgrades.

%
%

\smallskip

\noindent \textbf{Current and future work:} In the future we plan to obtain runtime results for systems of varying sizes. In particular, we are interested in real-world applications in the field of SPLs, together with other applications, such as modelling transition systems with access rights or deterioration.


On the more theoretical side of things, we have worked out the
coalgebraic concepts for CTSs \cite{bkks:cts-upgrades-coalgebraic} and
compared the matrix multiplication algorithm to the final chain
algorithm presented in \cite{ABHKMS12}, when applied to CTSs.

\smallskip

\noindent\textbf{Acknowledgements:} We thank Filippo
Bonchi and Mathias H\"ulsbusch for interesting discussions on
earlier drafts.

\bibliographystyle{plainurl}
\bibliography{references}

\full{
\appendices

\section{Proofs}
\label{sec:proofs}

Here we give proofs for all lemmas and propositions for which we have
omitted the proofs in the article.

\subsection{Proofs concerning Lattices and Lattice Transition Systems}
\label{sec:proofs-lattices}

\noindent\textbf{Proof of Lemma~\ref{lem:approximation}}

\begin{IEEEproof}
  Let $\ell,m\in\mathbb{B}$. Monotonicity of the approximation is
  immediate from the definition.

  \smallskip

  We next show
  $\app{\ell\sqcap m}\sqsupseteq \app{\ell}\sqcap\app{m}$: by
  definition we have $\app{\ell}\sqsubseteq \ell$,
  $\app{m}\sqsubseteq m$ and hence
  $\app{\ell}\sqcap\app{m}\sqsubseteq \ell\sqcap m$.  Since
  $\app{\ell\sqcap m}$ is the best approximation of $\ell\sqcap m$ and
  $\app{\ell}\sqcap\app{m}$ is one approximation, the inequality
  follows.

  In order to prove
  $\app{\ell\sqcap m}\sqsubseteq \app{\ell}\sqcap\app{m}$ observe that
  $\app{\ell}\sqsupseteq \app{\ell\sqcap m}$ and
  $\app{m}\sqsupseteq \app{\ell\sqcap m}$ by monotonicity of the
  approximation. Hence $\app{\ell\sqcap m}$ is a lower bound of
  $\app{\ell},\app{m}$, which implies
  $\app{\ell}\sqcap \app{m} \sqsupseteq \app{\ell}\sqcap\app{m}$.

  \smallskip

  Now let $\ell,m\in\mathbb{L}$. Recall the definitions
  $\app{\ell\sqcup\neg m}=\bigsqcup\{x\in\mathbb L\mid x\sqsubseteq
  \ell\sqcup\neg m\}$ and $m\rightarrow_\mathbb L
  \ell=\bigsqcup\{x\in\mathbb L\mid m\sqcap x\sqsubseteq \ell\}$. We
  will prove that both sets are equal.

  Assume $x\in\mathbb{L}$ with $x\sqsubseteq \ell\sqcup\neg m$, then
  $m\sqcap x\sqsubseteq m\sqcap(\ell\sqcup\neg m)=(m\sqcap \ell)\sqcup
  (m\sqcap\neg m) = (m\sqcap \ell)\sqcup 0 = m\sqcap \ell\sqsubseteq
  \ell$. For the other direction assume $m\sqcap x\sqsubseteq \ell$,
  then $\ell\sqcup\neg m\sqsupseteq (m\sqcap x)\sqcup\neg m = (m\sqcup
  \lnot m)\sqcap (x\sqcup \neg m) = 1\sqcap (x\sqcup \neg m) =
  x\sqcup\neg m\sqsupseteq x$.
\end{IEEEproof}

\noindent\textbf{Proof of Theorem~\ref{thm:CTSequiv}}

\begin{IEEEproof}
  Given a set $X$, a partially ordered set $(\Phi,\leq)$, and
  $\mathbb L=\mathcal O(\Phi)$, we define an isomorphism between the
  sets $(\Phi \xrightarrow{\text{mon.}} \mathcal{P}(X))^{X\times A}$
  and $ \mathcal O(\Phi)^{X \times A \times X}$. Consider the
  following function mappings
  $\eta: (\Phi \xrightarrow{\text{mon.}} \mathcal{P}(X))^{X\times
    A}\rightarrow \mathcal O(\Phi)^{X \times A \times X}, f\mapsto
  \eta(f)$
  and
  $\eta': \mathcal O(\Phi)^{X \times A \times X}\rightarrow(\Phi
  \xrightarrow{\text{mon.}} \mathcal{P}(X))^{X\times A}, \alpha
  \mapsto \eta'(\alpha)$ defined as:
  \begin{align*}
    \eta(f)(x,a,x')=&\ \{\phi \in \Phi \mid x' \in f_\phi(x,a) \},\\
    \eta'(\alpha)(x,a)(\phi)=&\ \{x' \mid \phi\in \alpha(x,a,x')\}.
  \end{align*}
  \begin{description}
  \item[Downward-closed] Let $\phi\in \eta(f)(x,a,x')$ and
    $\phi'\leq \phi$. By using these facts in the definition of
    $f_{\phi'}$ we find $x'\in f_{\phi'}(x,a)$, i.e.,
    $\phi' \in \eta(f)(x,a,x')$.
  \item[Anti-monotonicity] Let $\phi \leq \phi'$ and
    $x'\in \eta'(\alpha)(x,a)(\phi')$. Then by the above construction
    we find $\phi' \in \alpha(x,a,x')$. And by downward-closedness of
    $\alpha(x,a,x')$ we get $\phi\in\alpha(x,a,x')$, i.e.,
    $x'\in \eta'(\alpha)(x,a)(\phi)$.
  \end{description}
  Now it suffices to show that $\eta,\eta'$ are inverse of each other
  because by the uniqueness of inverses we then have
  $\eta'=\eta^{-1}$. We only give the proof of
  $\eta' \circ \eta = \mathsf{id}$, the proof of the other case
  ($\eta \circ \eta' = \mathsf{id}$) is similar. The former follows
  from the following observation:
  \begin{equation*}\begin{aligned} &x' \in f(x,a)(\phi) \Leftrightarrow x' \in f_\phi(x,a)
  \Leftrightarrow \phi \in \eta(f) (x,a,x') \\\Leftrightarrow& x'\in
  \eta'(\eta(f))(x,a)(\phi)\enspace.  \end{aligned}\end{equation*}
\end{IEEEproof}

\noindent\textbf{Proof of Theorem~\ref{thm:bisim-correspondence}}

\begin{IEEEproof}
  Let $x\in X,y\in Y$ be any two states in CTSs (LaTSs)
  $(X,A,f),(Y,A,g)$ over the conditions $\Phi$
  ($(X,A,\alpha),(Y,A,\beta)$ over the lattice $\mathcal O(\Phi\le)$),
  respectively.

  \noindent ($\Leftarrow$) Let $\phi \in \Phi$ be a condition and let
  $R$ be a lattice bisimulation relation such that $\phi \in R(x,y)$.
  Then, we can construct a family of relations $R_{\phi'}$ (for
  $\phi'\leq \phi$) as follows:
  $x R_{\phi'} y \Leftrightarrow \phi' \in R(x,y)$.  For all other
  $\phi'$ we set $R_{\phi'} = \emptyset$. The downward-closure of
  $R(x,y)$ ensures that $R_{\phi''} \subseteq R_{\phi'}$ (for
  $\phi',\phi'' \leq \phi$), whenever $\phi' \leq \phi''$.

  Thus, it remains to show that every relation $R_{\phi'}$ is a
  bisimulation. Let $x R_{\phi'} y$ and $x'\in f_{\phi'}(x,a)$. Then,
  $x \xrightarrow{a,\history{\phi'}} x'$. Since $\history{\phi'}$ is
  an irreducible in the lattice, $\history{\phi'} \subseteq R(x,y)$
  and $R$ is a lattice bisimulation, we find
  $y \xrightarrow{a,\history{\phi'}} y'$ and
  $\history{\phi'}\subseteq R(x',y')$, which implies
  $\phi'\in R(x',y')$. That is, $y'\in g_{\phi'} (y,a)$ and
  $x' R_{\phi'} y'$. Likewise, the remaining symmetric condition of
  bisimulation can be proved.

  \smallskip

  \noindent ($\Rightarrow$) Let $\sim_{\phi}$ be a conditional
  bisimulation between the CTSs $(X,A,f),(Y,A,g)$, for some
  $\phi \in \Phi$. Then, construct a conditional relation:
  $R(x,y) = \{\phi \mid x \sim_\phi y\}$. Clearly, the set $R(x,y)$ is
  a downward-closed subset of $\Phi$ due to
  Definition~\ref{def:cts-bisim}(ii); i.e., an element in the lattice
  $\mathcal O(\Phi)$. Next, we show that $R$ is a lattice
  bisimulation.

  Let $x \xrightarrow {a,\history {\phi'}} x'$ and
  $\history \phi' \subseteq R(x,y)$.  This implies
  $x' \in f_{\phi'}(x,a)$ and $\phi'\in R(x,y)$, hence
  $x \sim_{\phi'} y$. So using the transfer property of traditional
  bisimulation, we obtain $y'\in g_{\phi'}(y,a)$ and
  $x' \sim_{\phi'} y'$. That is, $y \xrightarrow{b,\history \phi'} y'$
  and $\phi'\in R(x',y')$, which implies
  $\history\phi' \subseteq R(x',y')$.  Likewise the symmetric
  condition of lattice bisimulation can be proved.
\end{IEEEproof}

\noindent\textbf{Proof of Lemma~\ref{lem:unionisbisim}}

\begin{IEEEproof}
  Let $x,x'\in X,a\in A,y\in Y$ and $\ell\in \mathcal J(\mathbb L)$
  such that $\ell \sqsubseteq \bigsqcup_{i\in I} R_i(x,y)$ and
  $x \xrightarrow{a,\ell} x'$. Then, there is an index $i\in I$ such
  that $\ell \sqsubseteq R_i(x,y)$, since $\ell$ is an irreducible.
  Thus, there is a $y'$ such that $y \xrightarrow{a,\ell} y'$ and
  $\ell \sqsubseteq R_i(x',y') \sqsubseteq \bigsqcup_{i\in I}
  R_i(x',y')$.
  Likewise, the symmetric condition when a transition emanates from
  $y$ can be proved.
\end{IEEEproof}

\medskip

\noindent\textbf{Proof of Theorem~\ref{thm:fixpoint-lattice-bisim}}
\begin{IEEEproof}
  \mbox{}

  \noindent ($\Leftarrow$) Let $R:X \times Y \rightarrow \mathbb L$ be
  a conditional relation over a pair of LaTSs
  $(X,A,\alpha),(Y,A,\beta)$ such that $R\sqsubseteq F(R)$. Next, we
  show that $R$ is a lattice bisimulation. For this purpose, let
  $\ell\in\irreducibles$, $a\in A$.  Furthermore, let
  $x\xrightarrow{a,\ell} x'$ (which implies
  $\ell \sqsubseteq \alpha(x,a,x')$) and $\ell\sqsubseteq R(x,y)$.
  From $R(x,y)\sqsubseteq F_1(R)(x,y)$ we infer
  $\ell\sqsubseteq F_1(R)(x,y)$. This means that
  $\ell \sqsubseteq \alpha(x,a,x') \rightarrow \big(\bigsqcup_{y'\in
    Y}(\beta(y,a,y')\sqcap R(x',y'))\big)$.
  Since $\ell_1\sqcap (\ell_1\to \ell_2) \sqsubseteq \ell_2$ we can
  take the infimum with $\alpha(x,a,x')$ on both sides and obtain
  $\ell \sqsubseteq \ell\sqcap \alpha(x,a,x') \sqsubseteq
  \bigsqcup_{y'\in Y}(\beta(y,a,y')\sqcap R(x',y'))$
  (the first inequality holds since
  $\ell \sqsubseteq \alpha(x,a,x')$). Since $\ell$ is irreducible
  there exists a $y'$ such that $\ell\sqsubseteq \beta(y,a,y')$,
  i.e., $y\xrightarrow{a,\ell} y'$, and $\ell\sqsubseteq R(x',y')$.



  Likewise, the remaining condition when a transition emanates from
  $y$ can be proved.

  \smallskip

  \noindent ($\Rightarrow$) Let $R:X\times Y \rightarrow \mathbb L$ be
  a lattice bisimulation on $(X,A,\alpha),(Y,A,\beta)$. Then, we need
  to show that $R\sqsubseteq F(R)$, i.e., $R\sqsubseteq F_1(R)$ and
  $R \sqsubseteq F_2(R)$. We will only give the proof of the former
  inequality, the proof of the latter is analogous. To show
  $R\sqsubseteq F_1(R)$, it is sufficient to prove
  $\ell\sqsubseteq R(x,y) \Rightarrow \ell\sqsubseteq F_1(R)(x,y)$,
  for all $x\in X,y\in Y$ and all irreducibles $\ell$. So let
  $\ell \sqsubseteq R(x,y)$, for some $x,y$.  Next, simplify $F_1(R)$
  as follows: {\allowdisplaybreaks
  \begin{align*}
    &F_1(R)(x,y)\\
    =&\bigsqcap_{a,x'} (\alpha(x,a,x') \rightarrow
                   \bigsqcup_{y'\in Y} (\beta (y,a,y') \sqcap R(x',y'))) \\
     =&\ \bigsqcap_{a,x'}\ \Big\lfloor\bigsqcup_{y'\in Y} (\beta (y,a,y')
        \sqcap R(x',y')) \sqcup \neg \alpha(x,a,x')\Big\rfloor \quad
        \mbox{(L.~\ref{lem:approximation})}\\
     =&\ \bigsqcap_{a,x'}\ \bigsqcup \{m\in \mathbb L \mid m
        \sqsubseteq \bigsqcup_{y'\in Y} (\beta (y,a,y') \sqcap
        R(x',y')) \\
        &\sqcup \neg \alpha(x,a,x')\}.
  \end{align*}
} Thus, it is sufficient to show that
$\ell \sqsubseteq \bigsqcup_{y'\in Y} (\beta (y,a,y') \sqcap R(x',y'))
\sqcup \neg \alpha(x,a,x')$,
for any $a\in A,x'\in X$. We do this by distinguishing the following
cases: either $\ell \sqsubseteq \neg \alpha(x,a,x')$ or
$\ell \sqsubseteq \alpha(x,a,x')$. If the former holds (which
corresponds to the case where there is no $a$-labelled transition under $\ell$),
then the result holds trivially. So assume
$\ell \sqsubseteq \alpha(x,a,x')$. Recall, from above, that
$\ell\sqsubseteq R(x,y)$ and $R$ is a lattice bisimulation. Thus,
there is a $y'\in Y$ such that $\ell \sqsubseteq \beta (y,a,y')$ and
$\ell \sqsubseteq R(x',y')$; hence,
  \[\ell \sqsubseteq \bigsqcup_{y'\in Y} (\beta (y,a,y') \sqcap
  R(x',y')) \sqcup \neg \alpha(x,a,x')\enspace. \]
\end{IEEEproof}

\noindent\textbf{Proof of Proposition~\ref{prop:lat-bisim&bool-bisim}}
\begin{IEEEproof}
  \mbox{}
  \begin{enumerate}[label=(\roman*)]
  \item This follows directly from Lemma~\ref{lem:approximation},
    allowing to move the approximations to the inside towards the
    implications and Lemma~\ref{lem:approximation}, allowing to
    approximate the implication in $\mathbb B$ via the implication in
    $\mathbb L$.
  \item If $R$ is a bisimulation in $\mathbb L$, then
    $F_\mathbb L(R)\sqsupseteq R$. Since by definition,
    $\app{Q}\sqsubseteq Q$ for all conditional relations $Q$ and we have
    shown in~(i) that $F_\mathbb L(R)=\app{F_\mathbb B(R)}$, we can
    conclude
    $F_\mathbb B(R)\sqsupseteq\app{F_\mathbb B(R)}=F_\mathbb
    L(R)\sqsupseteq R$.  Thus, $R$ is a $\mathbb B$-bisimulation.
  \item Clearly $R\sqsubseteq F_\mathbb B(R)$ because $R$ is a
    $\mathbb B$-bisimulation. Since $R$ has exclusively entries from
    $\mathbb L$, $F_\mathbb B(R) = \app{F_\mathbb B(R)}$, and finally~(i)
    yields that $\app{F_\mathbb B(R)} = F_\mathbb L (R)$; thus, $R$ is an
    $\mathbb L$-bisimulation.
  \end{enumerate}
\end{IEEEproof}

\noindent\textbf{Proof of Theorem~\ref{thm:fitting}}\label{app:proof:fitting}

\begin{IEEEproof}
  \mbox{}

  \noindent ($\Leftarrow$) Let $R:X\times X\rightarrow \mathbb B$ be a
  conditional relation satisfying
  $R\cdot\alpha \sqsubseteq \alpha \cdot R$ and
  $R^T \cdot \alpha \sqsubseteq \alpha \cdot R^T$. Then, we need to
  show that $R$ is a lattice bisimulation. Let $x \xrightarrow \ell y$
  such that $\ell \in \mathcal J (\mathbb B)$ and
  $\ell \sqsubseteq R(x,x')$. Then, we find
  $\ell \sqsubseteq \alpha(x ,y)$. That is,
  \begin{equation*}\begin{aligned}&\ell \sqsubseteq R(x,x') \sqcap \alpha(x,y)
  = R^T(x',x) \sqcap \alpha(x,y) \\\sqsubseteq&
  (R^T\cdot\alpha)(x',y) \sqsubseteq (\alpha\cdot R^T) (x',y)
  .\end{aligned}\end{equation*}
  By expanding the last term from above, we find that
  $\ell \sqsubseteq \alpha(x',y') \sqcap R^T(y',y)$, for some
  $y'$. Thus, $\ell \sqsubseteq \alpha(x',y')$ (which implies
  $x' \xrightarrow{\ell} y'$) and $\ell \sqsubseteq R(y,y')$.
  Similarly, the remaining condition when the transition emanates from
  $x'$ can be verified using
  $R \cdot \alpha\sqsubseteq \alpha \cdot R $.

  \smallskip

  \noindent ($\Rightarrow$) Let $R:X\times X\rightarrow \mathbb B$ be
  a lattice bisimulation. Then, we only prove
  $R\cdot\alpha \sqsubseteq \alpha\cdot R$; the proof of
  $R^T \cdot \alpha \sqsubseteq \alpha \cdot R^T$ is similar. Note
  that, for any $x,y'\in X$, we know that the element
  $(R\cdot\alpha)(x,y')$ can be decomposed into a set of atoms, since
  $\mathbb B$ is an atomic Boolean algebra. Let
  $(R\cdot \alpha)(x,y')=\bigsqcup_i \ell_i$ for some index set $I$
  such that the $\ell_i$ are atoms or irreducibles in $\mathbb B$.


  Furthermore, expanding the above inequality we get, for every
  $i\in I$ there is a state $y\in X$ such that
  $\ell_i \sqsubseteq R(x,y) \sqcap \alpha(y,y')$, since the
  $\ell_i$ are irreducibles. That is, for every $i\in I$ we have some
  state $y$ such that $\ell_i\sqsubseteq R(x,y)$ and
  $\ell_i \sqsubseteq \alpha(y,y')$. Now using the transfer property
  of $R$ we find some state $x'$ such that
  $\ell_i \sqsubseteq \alpha (x,x')$ and $\ell_i \sqsubseteq R(x',y')$.
  Thus, for every $i\in I$ we find that
  $\ell_i\sqsubseteq (\alpha\cdot R) (x,y')$; hence, since
  $(\alpha\cdot R)(x,y')$ is an upper bound of all $\ell_i$,
  $(R\cdot\alpha)(x,y')\sqsubseteq (\alpha\cdot R)(x,y')$.
\end{IEEEproof}

\subsection{Strategies for the Bisimulation Game}
\label{sec:proofs-strategies}

In the main text we claimed that there exists a winning strategy for
Player~2 in the conditional bisimulation game if and only if the start
states are conditionally bisimilar. In this section we will describe
the strategy and prove that it is correct.

\begin{lemma}
  Given two CTSs $(X,A,f)$, $(Y,A,g)$ and an instance $(x,y,\phi)$ of
  a bisimulation game, then whenever $x\sim_\phi y$, Player~2 has a
  winning strategy for $(x,y,\phi)$.
\end{lemma}

\begin{IEEEproof}
  The strategy of Player~2 can be directly derived from the family of
  CTS bisimulation relations $\{R_{\phi'}\mid\phi'\in\Phi\}$ where
  $(x,y)\in R_\phi$.  The strategy works inductively. Assume at any
  given point of time in the game, we have that the currently
  investigated condition is $\phi$ and $(x,y)\in R_{\phi}$, where $x$
  and $y$ are the currently marked states in $X$ respectively $Y$.
  Then Player~1 upgrades to $\phi'\leq\phi$. Due to the condition on
  CTS bisimulations of reverse inclusion, we have
  $R_{\phi'}\supseteq R_{\phi}$, therefore $(x,y)\in R_{\phi'}$.
  Then, when Player~1 makes a step $x\xrightarrow{a,\phi'}x'$ in $f$,
  there must exist a transition $y\xrightarrow{a,\phi'}y'$ in $g$ such
  that $(x',y')\in R_{\phi'}$ due to $R_{\phi'}$ being a (traditional)
  bisimulation. Analogously if Player~1 chooses a transition
  $y\xrightarrow{a,\phi'}y'$ in $g$, there exists a transition
  $x\xrightarrow{a,\phi'}x'$ in $f$ for Player~2 such that
  $(x',y')\in R_{\phi'}$. Hence, Player~2 will be able to react and
  establish the inductive condition again.  In the beginning, the
  condition holds per definition. Thus, Player~2 has a winning
  strategy.



\end{IEEEproof}



We will now prove the converse by explicitly constructing a winning
strategy for Player~1 whenever two states are not in a bisimulation
relation.


\begin{lemma}
  Given two CTSs $A,B$ and an instance $(x,y,\phi)$ of a bisimulation
  game, then whenever $x\not\sim_\phi y$, Player~1 has a winning
  strategy for $(x,y,\phi)$.
\end{lemma}

\begin{IEEEproof}
  We consider the LaTSs which correspond to the CTSs $A$, $B$ and
  compute the fixpoint by using the matrix multiplication algorithm,
  obtaining a sequence $R_0,R_1,\dots,R_n=R_{n+1}=\dots$ of
  lattice-valued relations
  $R_i\colon X\times Y\to \mathcal{O}(\Phi,\le)$. Note that instead of
  using exactly the matrix multiplication method, we can also use the
  characterisation of Definition~\ref{def:lattice-bisimulation}:
  whenever there exists a transition $x\xrightarrow{a,\phi} x'$, for
  which there is no matching transition with
  $y\xrightarrow{y,\phi} y'$ with $\phi\in R_{i-1}(x',y')$, the
  condition $\phi$ and all larger conditions $\phi'\ge \phi$ have to
  be removed from $R_{i-1}(x,y)$ in the construction of $R_i(x,y)$.

  We will now define
  $M^{\phi'}(x,y) = \max\{i\in \mathbb{N}_0\mid \phi'\in R_i(x,y)\}$,
  where $\max \mathbb{N}_0 = \infty$.  An entry
  $M^{\phi'}(x,y) = \infty$ signifies that $x\sim_{\phi'} y$, whereas
  any other entry $i < \infty$ means that $x,y$ were separated under
  condition $\phi'$ at step $i$ and hence $x\not\sim_{\phi'} y$.



  Now assume we are in a game situation with game instance
  $(x,y,\phi)$ where Player~1 has to make a step. We will show that if
  $M^\phi(x,y)=i < \infty$, Player~1 can choose an upgrade
  $\overline{\phi}\le \phi$, an action $a\in A$ and a step
  $x\xrightarrow{a,\overline \phi}x'$ (or
  $y\xrightarrow{a,\overline \phi}y'$) such that independently of the
  choice of the corresponding state $y'$, respectively $x'$, which
  Player~2 makes, $M^{\overline \phi}(x',y') < i$.
	
  For each $\phi'\le \phi$ compute
  \begin{equation*}
    \begin{aligned}
      &\omega(\phi')\\
      =&\min&\{\min_{a,x'}\{\max_{y'}\{M_n^{\phi'}(x',y')
      \mid y\xrightarrow{a,\phi'}y'\}\mid x\xrightarrow{a,\phi'}x'\},\\
      &&\{\min_{a,y'}\{\max_{x'}\{M_n^{\phi'}(x',y')\mid
      x\xrightarrow{a,\phi'}x'\}\mid
      y\xrightarrow{a,\phi'}y'\}\}
    \end{aligned}
  \end{equation*}

  The formula can be interpreted as follows: The outer $\min$
  corresponds to the choice of making a step in transition system $A$
  or $B$. The inner $\min$ corresponds to choosing the step that
  yields the best, i.e. lowest, guaranteed separation value and the
  $\max$ corresponds to the choice of Player 2 that yields the best,
  i.e. greatest, separation value for him.

  Now choose a minimal condition $\overline{\phi}$ such that
  $\omega(\overline{\phi})$ is minimal for all
  $\phi'\le \phi$.
  Player~1 now makes an upgrade
  from $\phi$ to $\overline{\phi}$ and chooses a transition
  $x\xrightarrow{a,\overline{\phi}} x'$ or
  $y\xrightarrow{a,\overline{\phi}} y'$ such that the minimum in
  $\omega(\overline{\phi})$ is reached. This means that Player~2 can
  only choose a corresponding successor state $y'$ respectively $x'$
  such that $M^{\overline{\phi}}(x',y') \le \omega(\overline{\phi})$.

  Now it remains to be shown that $\omega(\overline{\phi}) < i$, via
  contradiction: assume that $\omega(\overline{\phi}) \ge i$. Since
  $\omega(\overline{\phi})$ is minimal for all $\phi'\le \phi$, we
  obtain $\omega(\phi') \ge i$ for all $\phi'\ge \phi$. This implies
  that for each step $x\xrightarrow{a,\phi'} x'$ there exists an
  answering step $y\xrightarrow{a,\phi'} y'$ such that
  $M^{\phi'}(x',y') \ge i$ (analogously for every step of $y$).  The
  condition $M^{\phi'}(x',y') \ge i$ is equivalent to
  $\phi'\in R_i(x',y')$ and hence we can infer that
  $\phi'\in R_{i+1}(x,y)$. This also holds for $\phi'=\phi$, which is
  a contradiction to $M^{\phi}(x,y) = i$.

	
  In order to conclude, take two states $x,y$ and a condition $\phi$
  such that $x\not\sim_\phi y$. Then $M^\phi(x,y) = i < \infty$ and
  the above strategy allows Player~1 to force Player~2 into a game
  instance $(x',y',\overline{\phi})$ where
  $M^{\overline{\phi}}(x',y') < M^\phi(x,y)$. Whenever
  $M^\phi(x,y) = 1$ Player~1 wins immediately, because then $x$ allows
  a transition that $y$ can not mimic or vice-versa, and Player~1
  simply takes this transition. Therefore we have found a winning
  strategy for Player~1.

\end{IEEEproof}

\subsection{Proofs concerning ROBDDs}

\noindent\textbf{Proof of Lemma~\ref{lem:highsmallerlow}}

\begin{IEEEproof}
  \mbox{}

  \noindent ($\Rightarrow$) Assume that
  $\mathit{low}(n)\models \mathit{high}(n)$ for all nodes $n$ of $b$.

  Let $C'\in\llbracket b\rrbracket$ and $C\le C'$. Without loss of
  generality we can assume that $C = C'\cup\{f\}$ for some $f\in U$.
  (The rest follows from transitivity.) For the configuration $C'$
  there exists a path in $b$ that leads to $1$. We distinguish the
  following two cases:
  \begin{itemize}
  \item There is no $f$-labelled node on the path. Then the path for
    $C$ also leads to $1$ and we have $C\in\llbracket b\rrbracket$.
  \item If there is an $f$-labelled node $n$ on the path, then $C'$
    takes the \emph{low}-successor, $C$ the \emph{high}-successor of
    this node. Since $\mathit{low}(n)\models \mathit{high}(n)$ we
    obtain
    $\llbracket\mathit{low}(n)\rrbracket \subseteq
    \llbracket\mathit{high}(n)\rrbracket$.
    Hence the remaining path for $C$, which contains the same features
    as the path for $C'$, will also reach $1$.
  \end{itemize}

  \smallskip

  \noindent ($\Leftarrow$) Assume by contradiction that
  $\llbracket b\rrbracket$ is downward-closed, but there exists a node
  $n$ with $\mathit{low}(n)\not\models \mathit{high}(n)$ and
  $f = \mathit{root}(n) \in U$. Hence there must be a path from the
  \emph{low}-successor that reaches $1$, but does not reach $1$ from
  the \emph{high}-successor. Prefix this with the path that reaches
  $n$ from the root of $b$.

  In this way we obtain two configurations $C = C'\cup\{f\}$, i.e.,
  $C\le C'$, where $C'\in\llbracket b\rrbracket$, but
  $C\not\in\llbracket b\rrbracket$. This is a contradiction to the
  fact that $\llbracket b\rrbracket$ is downward-closed.

\end{IEEEproof}

\noindent\textbf{Proof of Lemma~\ref{lem:approximationROBDD}}

\begin{IEEEproof}
  \mbox{}

  \begin{itemize}
  \item We show that $\approxL{b}$ as obtained by
    Algorithm~\ref{alg:approxROBDD} is downward-closed. This can be seen via
    induction over the number of different features occurring in the
    BDD~$b$. If $b$ only consists of a leaf node, then $\approxL{b}$
    is certainly downward-closed.  Otherwise, we know from the
    induction hypothesis that $\approxL{\mathit{high}(b)}$,
    $\approxL{\mathit{low}(b)}$ are downward-closed.  If
    $\mathit{root}(b)\not\in U$, then $\approxL{b}$ is downward-closed
    due to Lemma~\ref{lem:highsmallerlow}. If however
    $\mathit{root}(b)\in U$, then
    $\approxL{\mathit{high}(b)}\land \approxL{\mathit{low}(b)}$ is
    downward-closed (since downward-closed sets are closed under
    intersection). Furthermore
    $\approxL{\mathit{high}(b)}\land \approxL{\mathit{low}(b)} \models
    \approxL{\mathit{high}(b)}$,
    i.e., the new \emph{low}-successor implies the
    \emph{high}-successor. That means that the condition of
    Lemma~\ref{lem:highsmallerlow} is satisfied at the root and elsewhere
    in the BDD and hence the resulting BDD $\approxL{b}$ is
    downward-closed.

  \item First, from the construction where a \emph{low}-successor is
    always replaced by a stronger \emph{low}-successor, it is easy to
    see that $\approxL{b}\models b$.

    We now show that there is no other downward-closed ROBDD $b'$ such
    that $\approxL{b}\models b'\models b$: Assume to the contrary that there
    exists such a downward-closed BDD $b'$. Hence there exists a
    configuration $C\subseteq N$ such that $C\not\models \approxL{b}$,
    $C\models b'$, $C\models b$. Choose $C$ maximal wrt.\ inclusion.

    Now we show that there exists a feature $f\in U$ such that
    $f\not\in C$ and $C\cup\{f\} = C'\not\models b$. If this is the
    case, then $C'\le C$ and $C'\not\models b'$, which is a
    contradiction to the fact that $b'$ is downward-closed.

    Consider the sequence $b=b_0,\dots,b_m=\approxL{b}$ of BDDs that is
    constructed by the approximation algorithm
    (Algorithm~\ref{alg:approxROBDD}), where the BDD structure is upgraded
    bottom-up. We have
    $\approxL{b}=b_m\models b_{m-1}\models \dots \models b_0=b$, since in each
    newly constructed BDD for some node $n$ $\mathit{low}(n)$ with
    $\mathit{root}(n)\in U$ is replaced by
    $\mathit{high}(n)\land \mathit{low}(n)$.

    Since $C\models b$ and $C\not\models \approxL{b}$, there must be an index
    $k$ such that $C\models b_k$, $C\not\models b_{k+1}$. Let $n$ be
    the node that is modified in step~$k$, where
    $\mathit{root}(n) = f\in U$. We must have $f\not\in C$, since the
    changes concern only the \emph{low}-successor and if $f\in C$, the
    corresponding path would take the \emph{high}-successor and
    nothing would change concerning acceptance of $C$ from $b_k$ to
    $b_{k+1}$.

    Now assume that $C' = C\cup\{f\}\models b$. This would be a
    contradiction to the maximality of $C$ and hence
    $C\cup\{f\}\not\models b$, as required.

  \end{itemize}
\end{IEEEproof}

\noindent\textbf{Proof of Proposition~\ref{prop:residuuminOd}}

\begin{IEEEproof}
  \mbox{}
  \begin{enumerate}[label=(\roman*)]
  \item For this proof, we work with the set-based interpretation,
    which allows for four views, one on the Boolean algebra
    $\mathbb B = \mathcal P(\mathcal P(N))$, one on the lattice
    $\mathbb L = \mathcal O(\mathcal P(N),\leq)$, one of the Boolean
    algebra $\mathbb B'=\mathcal P(\llbracket d\rrbracket)$ and one on
    the lattice
    $\mathbb L'=(\mathcal O(\llbracket d\rrbracket), \leq')$ where
    $\leq'\,=\,\leq|_{\llbracket d\rrbracket\times \llbracket
      d\rrbracket}$.
    We will mostly argue in the Boolean algebra $\mathbb B$. When talking about
    downward-closed sets, we will usually indicate with respect to
    which order. Similarly, the approximation relative to $\leq$ is
    written $\app{\_}$, whereas the approximation relative to $\leq'$ is
    written $\app{\_}'$.

    We can compute:
    \begin{equation*}
      b_1\rightarrow_{\mathbb L'}b_2\equiv\app{\neg_{\mathbb B'}b_1\lor
      b_2}' \equiv \app{(\neg_\mathbb Bb_1\land d)\lor b_2}'
    \end{equation*}
    To conclude the proof, we will now show that
    $\app{b}' \equiv \app{b\lor\neg d}\land d$ for any $b\in\mathbb{B}'$. We
    prove this via mutual implication.
    \begin{itemize}
    \item We show $\app{b\lor \lnot d}\land d\models \app{b}'$:

      $$\app{b\lor\neg d} \land d \models (b\lor\neg d)\land d \equiv
      (b\land d)\lor(\neg d\land d) \equiv b\land d\models b$$

      Since $\app{b\lor\lnot d}\land d$ implies $d$, it certainly is in
      $\mathbb B'$. We now show that it is downward-closed wrt.\
      $\le'$: we use an auxiliary relation $\leq''$, which is the smallest
      partial order on $\mathbb{B}$ that contains $\le'$, i.e., $\le'$
      extended to $\mathbb{B}$. We have $\le''\,\subseteq\,\le$. Since
      $\app{b\lor\lnot d}$ is an approximation it is downward-closed wrt.\
      $\le$ and hence downward-closed wrt.\ $\le'$. Moreover, $d$ is
      downward-closed relative to $\leq''$ (obvious by definition).
      Since the intersection of two downward-closed sets is again
      downwards closed, $\app{b\lor \lnot d}\land d$ is downward-closed
      relative to $\leq''$ and since finally, downward-closure
      relative to $\leq''$ is the same as downward-closure relative to
      $\leq'$ provided we discuss an element from $\mathbb B'$, we can
      conclude that $\app{b\lor\lnot d}\land d$ belongs to $\mathbb L'$.





      From $\app{b\lor\lnot d}\land d\in\mathbb L'$ and
      $\app{b\lor\lnot d}\land d\models b$ it follows that
      $\app{b\lor\lnot d}\land d\models \app{b}'$ by definition of the
      approximation.
    \item We show $\app{b}'\models \app{b\lor \lnot d}\land d$:
				
      Let any $C\in\mathcal P(N)$ be given, such that
      $C\in \llbracket \app{b}'\rrbracket$. We show that in this case
      $C \in \llbracket \app{b\lor \lnot d}\land d\rrbracket$, which
      proves $\app{b}'\models \app{b\lor \lnot d}\land d$. Let $\history C$
      be the downwards-closure of $C$ wrt.\ $\le$.
				
      Since $\app{b}'$ must de downward-closed relative to $\leq'$, it
      holds that
      $\history C \cap \llbracket d\rrbracket \subseteq \llbracket
      \app{b}'\rrbracket$.
      Disjunction with $\lnot d$ on both sides yields
      $\history C \subseteq \llbracket \app{b}'\lor\lnot d\rrbracket
      \subseteq \llbracket b\lor\lnot d\rrbracket$,
      since $c\models c\lor\lnot d\equiv (c\land d)\lor \lnot d$.  The
      set $\history C$ is downwards-closed wrt.\ $\le$, so it is
      contained in the approximation relative to $\leq$ of this set,
      i.e $\history C\subseteq \llbracket \app{b\lor\lnot d}\rrbracket$.
      Thus, in particular,
      $C \in \llbracket \app{b\lor\lnot d}\rrbracket$. Since
      $C \in \llbracket\app{b}'\rrbracket$, it follows that
      $C\in \llbracket d\rrbracket$, therefore we can conclude
      $C\in \llbracket \app{b\lor\lnot d}\land d\rrbracket$.
				
    \end{itemize}
    Hence
    \begin{equation*}
      \begin{aligned}
        & \app{(\neg_\mathbb Bb_1\land d)\lor b_2}' \equiv \app{(\neg_\mathbb
        Bb_1\land d)\lor b_2 \lor \lnot d} \land d \\
        \equiv&  \app{\neg_\mathbb
        Bb_1\lor b_2 \lor \lnot d} \land d
      \end{aligned}
    \end{equation*}
  \item Since $d$ is downward-closed wrt.\ $\le$, $d=\app{d}$, therefore,
    using Lemma~\ref{lem:approximation}, we obtain
    $\app{\lnot b_1\lor b_2\lor \lnot d}\land d\equiv \app{\lnot b_1\lor
    b_2\lor \lnot d}\land \app{d}\equiv \app{(\lnot b_1\lor b_2\lor \lnot
    d)\land d} \equiv \app{(\lnot b_1\lor b_2)\land d\lor \lnot d\land
    d} \equiv \app{(\lnot b_1\lor b_2)\land d}\equiv \app{\lnot
    b_1\lor b_2} \land
    \app{d}\equiv\app{\lnot b_1\lor b_2}\land d$.
  \end{enumerate}
\end{IEEEproof}

\section{Run-Time Results}
\label{sec:runtime-results}

In this section we present the detailed run-time results for our
BDD-based implementation versus the non-BDD-based implementation for a
sequence of CTSs.

Table~\ref{tab:run-time-results} shows the runtime results (in
milliseconds) for the computation of the largest bisimulation for our
implementation on the family of CTSs described in
Section~\ref{sec:impl-runtime}. Despite some fluctuations, the
quotient of the time taken when not using BDDs and when using BDDs
increases exponentially by factor of about~$2$.

\begin{figure*}
\begin{center}
  \begin{tabular}{|r|r|r|r|}\hline
    \# features & time(BDD) & time(without BDD) &
    time(without BDD)/time(BDD) \\ \hline
    1&42&13&0.3 \\ \hline
    2&64&32&0.5 \\ \hline
    3&143&90&0.6\\ \hline
    4&311&312&1.0 \\ \hline
    5&552&1128&2.0 \\ \hline
    6&1140&3242&2.8 \\ \hline
    7&1894&8792&4.6 \\ \hline
    8&1513&13256&8.8 \\ \hline
    9&1872&39784&21 \\ \hline
    10&3208&168178&52 \\ \hline
    11&5501&513356&93 \\ \hline
    12&7535&1383752&184 \\ \hline
    13&5637&3329418&591 \\ \hline
    14&6955&8208349&1180 \\ \hline
    15&11719&23700878&2022 \\ \hline
    16&15601&57959962&3715 \\ \hline
    17&18226&150677674&8267 \\ \hline
    18&17001&347281057&20427 \\\hline
    19&22145&out of memory&--- \\\hline
  \end{tabular}
  \caption{Run-time results (in milliseconds)}
  \label{tab:run-time-results}
\end{center}
\end{figure*}



\section{Deactivating Transitions}
\label{sec:deactivating-transitions}

We will now work on an extension that allows transitions to deactivate
when upgrading.


We have introduced conditional transition systems (CTS) as a modelling
technique that can be used for modelling software product lines
(SPLs). CTSs are a strictly stronger model than (standard) FTSs,
allowing for upgrades. Products derived from a software product line
may be upgraded to advanced versions, activating additional
transitions in the system. A change in the transition function can
only be realised in one direction: by adding transitions which were
previously not available, while all previously active transitions
remain active.

However, this choice may not be the optimal choice in all cases,
because sometimes an advanced version of a system may offer improved
transitions over the base product. For instance, a free version of a
system may display a commercial when choosing a certain transition,
whereas a premium model may forego the commercial and offer the base
functionality right away.

A practical motivation may be derived from our
Example~\ref{ex:cts-bisim}.  In this transition system one may want to
be able to model that in the unsafe state, the advanced version can
only send an encrypted message, since we assume that the user is
always interested in a secure communication, ensured either by a safe
channel or by encryption.  However, it is not an option to simply drop
the unencrpyted transition from the unsafe state with respect to the
base version, because then, whenever the system encounters an unsafe
state in the base version, the system will remain in a deadlock unless
the user decides to perform an upgrade. We will solve such a situation
as follows: we will add priorities that allow to deactivate the
unencrypted transition in the presence of an encrypted transition.

In order to allow for deactivation of transitions when upgrading, we
propose a slight variation of of the definition of CTS/LaTS and the
corresponding bisimulation relation. A \emph{conditional transition
  system with action precedence} is a triple $(X,(A,<_A),f)$, where
$(X,A,f)$ is a CTS and $<_A$ is a strict order on $A$.

Intuitively, a CTS with action precedence evolves in a very
similar way  to standard CTS. Before the system starts acting, it is
assumed that all the conditions are fixed and a condition
$\phi\in \Phi$ is chosen arbitrarily which represents a selection of a
valid product of the system (product line). Now all the transitions
that have a condition greater than or equal to $\phi$ are activated,
while the remaining transitions are inactive. This is unchanged from
standard CTS, however, \emph{if from a state $x$ there exist two
  transitions $x\xrightarrow{a,\phi}x'$ and
  $x\xrightarrow{a',\phi}x''$ where $a'>a$, i.e. $a'$ takes precedence
  over $a$, then additionally $x\xrightarrow{a,\phi}x'$ remains
  inactive}. Henceforth, the system behaves like a standard transition
system; until at any point in the computation, the condition is
changed to a smaller one (say, $\phi'$) signifying a selection of a
valid, upgraded product. Now (de)activation of transitions depends
on the new condition $\phi'$, rather than on the old condition $\phi$.
As before, active transitions remain active during an upgrade,
\emph{unless new active transitions appear that are exiting the same
  state and are labelled with an action of higher priority}.

In the sequel we will just write CTS for CTS with action precedence,
since for the remainder of this section we will solely investigate
this variation of CTS.  This changed interpretation of the behaviour
of a CTS of course also has an effect on the bisimulation.

\begin{defn}
  \label{def:cts-bisim-precedence}
  Let $(X,(A,<_A),f)$, $(Y,(A,<_A),g)$ be two CTSs over the same set
  of conditions $(\Phi,\leq)$. For a condition $\phi\in\Phi$, we
  define $\bar{f}_\phi(x,a)$ to denote the \emph{labelled transition
    system} induced by a CTS $(X,(A,<_A),f)$ with action precedence,
  where
  \begin{equation*}\begin{aligned} &\bar{f}_\phi(x,a) = \\
  & \{x' \mid
  x\xrightarrow{a,\phi}x'\wedge\neg(\exists a'\in A, x''\in X:
  a'>a\wedge x\xrightarrow{a',\phi}x'')\}.
\end{aligned}\end{equation*}
Two states $x\in X,y\in Y$ are \emph{conditionally bisimilar (wrt.\
  action precedence)} under a condition $\phi\in \Phi$, denoted
$x \sim^p_\phi y$, if there is a family of relations $R_{\phi'}$ (for
every $\phi'\leq \phi$) such that
\begin{enumerate}[label=(\roman*)]
\item each relation $R_{\phi'}$ is a traditional bisimulation relation
  between $\bar{f}_{\phi'}$ and $\bar{g}_{\phi'}$,
\item whenever $\phi'\leq \phi''$, we have $R_{\phi'}\supseteq R_{\phi''}$, and
\item $R_\phi$ relates $x$ and $y$, i.e. $(x,y)\in R_\phi$.
\end{enumerate}
\end{defn}

The definition of bisimilarity is analogous to traditional CTS but
refers to the new transition system given by $\bar{f}$, which contains
only the maximal transitions.

Lattice transition systems (LaTSs) can be extended in the same way, by
adding an order on the set of actions and leaving the remaining
definition unchanged. Disregarding deactivation, there still is a
duality between CTSs and LaTSs. Now, in order to characterise
bisimulation using a fixpoint operator, we can modify the operators
$F_1, F_2$ and $F$ to obtain $G_1$, $G_2$ and $G$, respecting the
deactivation of transitions as follows.

\begin{defn}
  \label{def:G-operator}
  Let $(X,(A,<_A),\alpha)$ and $(Y,(A,<_A),\beta)$ be LaTSs (with
  ordered actions). Recall the residuum operator ($\to$) on a lattice
  and define three operators
  $G_1,G_2,G:(X\times Y\rightarrow\mathbb L)\rightarrow(X\times
  Y\rightarrow\mathbb L)$ in the following way:

{
\allowdisplaybreaks
{\footnotesize
\begin{align*}
  G_1(R)(x,y) = \bigsqcap_{a\in A, x'\in X}\bigg(&\alpha(x,a,x') \rightarrow
                \big(\bigsqcup_{y'\in Y}(\beta(y,a,y')\sqcap
                R(x',y')) \\
  & \qquad \sqcup \bigsqcup_{a'>a, x''\in
                X}\alpha(x,a',x'')\big)\bigg), \\
  G_2(R)(x,y) = \bigsqcap_{a\in A,y'\in Y}\bigg(&\beta(y,a,y') \rightarrow
                \big(\bigsqcup_{x'\in X}(\alpha(x,a,x')\sqcap
                R(x',y'))  \\
  & \qquad \sqcup \bigsqcup_{a'>a, y''\in Y}\beta(y,a',y'')\big)\bigg),
\end{align*}
$$G(R)(x,y) =\ G_1(R)(x,y) \sqcap G_2(R)(x,y).$$
}
}
\end{defn}
Now, we need to show that we can characterise the new notion of
bisimulations as post-fixpoints of this operator $G$. For the
corresponding proof we will make use of the following observation:
\begin{lemma}
  \label{lem:approximpli}
  Let $\mathbb L=\mathcal O(\Phi)$ for any finite partially ordered
  set $(\Phi,\leq)$ be a lattice that embeds into
  $\mathbb B=\mathcal P(\Phi)$. Take $\phi\in\Phi$. Then, in order to
  show that $\phi\in (l_1\rightarrow l_2)$, for any given
  $l_1, l_2\in\mathbb L$, it suffices to show that for all
  $\phi'\leq\phi$, $\phi'\notin l_1$ or $\phi'\in l_2$.
\end{lemma}

\begin{IEEEproof}
  We have already shown that
  $l_1\rightarrow l_2=\lfloor l_1\rightarrow l_2\rfloor=\lfloor \neg
  l_1\sqcup l_2\rfloor$.
  Now, if all $\phi'\leq\phi$ are not in $l_1$, i.e. in $\neg l_1$, or
  in $l_2$, then all $\phi'\leq\phi$ are in $\neg l_1\sqcup l_2$.
  Therefore, $\downarrow\phi\subseteq \neg l_1\sqcup l_2$, and thus
  $\phi\in l_1\rightarrow l_2$.
\end{IEEEproof}

\begin{theorem}
  \label{thm:fixpoint-lattice-bisim-precedence}
  Let $(X,(A, <_A),f)$, $(Y,(A, <_A),g)$ be two CTSs over $(\Phi,\le)$
  and $(X,(A,<_A),\alpha)$, $(Y,(A,<_A),\beta)$ over
  $\mathcal O(\Phi)$ be the corresponding LaTS. For any two states
  $x\in X, y\in Y$ it holds that $x\sim_\phi^p y$ if and only if there
  exists a post-fixpoint $R:X\times Y\rightarrow\mathbb L$ of $G$
  ($R\sqsubseteq G(R)$) such that $\phi\in R(x,y)$.
\end{theorem}
\begin{IEEEproof}
  \mbox{}
  \begin{itemize}
  \item Assume $R$ is a post-fixpoint of $G$, i.e.
    $R\sqsubseteq G(R)$, let $x\in X$ and $y\in Y$ be given
    arbitrarily and $\phi\in R(x,y)$. We define for each
    $\phi'\leq\phi$ a relation $R_{\phi'}$ according to
    $$(x',y')\in R_{\phi'}\Leftrightarrow\phi'\in R(x',y').$$

    Since each set $R(x',y')$ is downward-closed for all
    $x'\in X, y'\in Y$, it holds that $R_{\phi_1}\subseteq R_{\phi_2}$
    whenever $\phi_1\geq\phi_2$. Moreover, since we assume
    $\phi\in R(x,y)$, $(x,y)\in R_{\phi'}$ must hold for all
    $\phi'\leq\phi$.

    So we only need to show that all $R_{\phi'}$ are traditional
    bisimulations for $\bar{f}_{\phi'}$.  For this purpose let
    $x', y',\phi'$ be given, such that $(x',y')\in R_{\phi'}$.
    Moreover, let $a\in A$ and $x''\in X$ be given such that
    $x''\in \bar{f}_{\phi'}(x',a)$ -- if no such $a$ and $x''$ exists
    then the first bisimulation condition is trivially true. For
    $G_1$, it must be true that $\phi'\in G_1(R)(x,y)$. Thus,
    \begin{align*}
      \phi'\in \bigg(\alpha(x',a,x'') \rightarrow &\
    \big(\bigsqcup_{y''\in Y}(\beta(y',a,y'')\sqcap R(x'',y''))
    \sqcup\\
    &\ \bigsqcup_{a'>a, x'''\in X}\alpha(x',a',x''')\big)\bigg)
    \end{align*}

    Since we also know that $\phi'\in\alpha(x',a,x'')$ because
    $x''\in \bar{f}_{\phi'}(x',a)$, it must be true that
    \begin{equation*}\begin{aligned}\phi'\in \bigg(&\bigsqcup_{y''\in Y}(\beta(y',a,y'')\sqcap
    R(x'',y'')) \\&\sqcup\bigsqcup_{a'>a, x'''\in
      X}\alpha(x',a',x''')\bigg).\end{aligned}\end{equation*}
    This is true because
    $\psi\in l_1\rightarrow l_2 \Leftrightarrow \psi\in\lfloor\neg
    l_1\vee l_2\rfloor\Rightarrow\psi\in \neg l_1\vee l_2$
    (Lemma~\ref{lem:approximpli}) and, if $\psi\in l_1$, hence
    $\psi\notin\neg l_1$, it follows that $\psi\in l_2$.
	
    Per definition of $\bar{f}_{\phi'}$, there exists no $a'>a$ such
    that $\bar{f}_{\phi'}(x'',a')\neq\emptyset$. Therefore,
    $$\phi'\notin\bigsqcup_{a'>a, x'''\in
      X}\alpha(x',a',x''').$$It follows that
    $$\phi'\in\bigsqcup_{y''\in Y}\beta(y',a,y'')\sqcap
    R(x'',y'').$$

    Then, there must exist at least one $y''\in Y$ such that
    $\phi'\in\beta(y',a,y'')\sqcap R(x'',y'')$. It follows that
    $\phi'\in R(x'',y'')$, i.e. $(x'',y'')\in R_{\phi'}$.
		
    We will now show that $y''\in \bar{g}_{\phi'}(y',a)$, holds as
    well.  Assume, to the contrary, that
    $y''\notin \bar{g}_{\phi'}(y',a)$, then, due to
    $\phi'\in\beta(y',a,y'')$, there must exist an $a'>a$ and a
    $y'''\in Y$ such that $\phi'\in\beta(y',a',y''')$. W.l.o.g.
    choose $a'$ maximal.  Since we required $(x',y')\in R_{\phi'}$, it
    has to hold that $\phi'\in G_2(R)(x',y')$. So in particular,
    \begin{align*}
      \phi'\in \bigg(&\beta(y',a',y''')
    \rightarrow\ \big(\bigsqcup_{x'''\in X}(\alpha(x',a',x''')\\
    &\sqcap
    R(x''',y''')) \ \sqcup\bigsqcup_{a''>a', y''''\in
      Y}\beta(y',a'',y'''')\big)\bigg)
    \end{align*}
    Since we chose $a'$ maximal, we know that
    $\phi'\notin\bigsqcup_{a''>a', y''''\in Y}\beta(y',a'',y'''')$.
    Moreover, since $a'>a$ and $x''\in \bar{f}_{\phi'}(x',a)$, there
    exists no $x'''$ such that $\phi'\in\alpha(x',a',x''')$.  Thus,
    $\phi'$ is not in the right side of the residuum, yet it is in the
    left side of the residuum, therefore, it is not in the residuum.
    Thus, we can conclude $\phi'\notin G_2(R)(x',y')$, which is a
    contradiction.
		
    Thus, the first bisimulation condition is true. The second
    condition can be proven analogously, reversing the roles of $G_2$
    and $G_1$ to find the answer step in $\bar{f}_{\phi'}$.
  \item Now, assume the other way around, that a family $R_\phi$ of
    bisimulations from $\bar{f}_\phi$ to $\bar{g}_\phi$ exists such
    that for all states $x\in X$, $y\in Y$ and for all pairs of
    conditions $\phi_1, \phi_2\in\Phi$ the expression
    $\phi_1\leq\phi_2$ implies $R_{\phi_1}\supseteq R_{\phi_2}$.
    Moreover, let $\phi$, $x\in X$ and $y\in Y$ be given such that
    $(x,y)\in R_\phi$. We define $R:X\times Y\rightarrow\mathbb L$
    according to
    $$R(x,y)=\{\phi'\mid (x,y)\in R_{\phi'}\}.$$Due to
    anti-monotonicity of the family of $R_{\phi'}$ all entries in $R$
    are indeed lattice elements from $\mathcal O(\Phi, \leq)$.
    Moreover, by definition, $\phi\in R(x,y)$. So it only remains to
    be shown that $R$ is a post-fixpoint.
	
    For this purpose, let $x'\in X$, $y'\in Y$ and $\phi'\in \Phi$ be
    given, such that $\phi'\in R(x',y')$. (If no such $x', y', \phi'$
    exist, then $R$ is the zero matrix (where all entries are
    $\emptyset$) and $R\sqsubseteq G(R)$ holds trivially.) We will now
    show that $\phi'\in G_1(R)(x',y')$. The fact that
    $\phi'\in G_2(R)(x',y')$ can be shown analogously. We need to show that
    \begin{equation*}\begin{aligned}\phi'\in \bigg(&\alpha(x,a,x') \rightarrow\big(\bigsqcup_{y'\in
      Y}(\beta(y,a,y')\sqcap R(x',y')) \\&\sqcup\bigsqcup_{a'>a, x''\in
      X}\alpha(x,a',x'')\big)\bigg)\end{aligned}\end{equation*} for all $x''\in X$ and $a\in A$.
	
    We recall that
    $l_1\rightarrow_\mathbb L l_2=\lfloor l_1\rightarrow_\mathbb B
    l_2\rfloor=\lfloor\neg l_1\vee l_2\rfloor$
    (Lemma~\ref{lem:approximation}) and show that whenever
    $\phi'\in\alpha(x,a,x')$, it holds that
    $\phi'\in\big(\bigsqcup_{y'\in Y}(\beta(y,a,y')\sqcap R(x',y'))
    \sqcup\bigsqcup_{a'>a, x''\in X}\alpha(x,a',x'')\big)$.
    We distinguish according to whether $a$ is maximal such that
    $\phi'\in\alpha(x,a,x'')$:
    \begin{itemize}
    \item There is no $a'>a$ such that $\phi'\in\alpha(x,a,x'')$ for
      any $x''\in X$:

      Then there must exist a $y'\in Y$ such that
      $\phi'\in\beta(y,a,y')$ and $(x',y')\in R_{\phi'}$, i.e.
      $\phi'\in R(x',y')$, because $R_{\phi'}$ is a bisimulation and
      for all $\phi''\leq\phi'$ we have
      $R_{\phi'}\subseteq R_{\phi''}$.
    \item There is an $a'>a$ such that $\phi'\in\alpha(x,a,x'')$ for
      some $x''\in X$:

      Then $\phi'\in\bigsqcup_{a'>a, x''\in X}\alpha(x,a',x'')$.
    \end{itemize}
    So we have shown for all $\phi'\in R(x',y')$ that
    $\phi'\in\alpha(x,a,x')$ implies
    \begin{align*}
      \phi'\in &\big(\bigsqcup_{y'\in Y}(\beta(y,a,y')\sqcap R(x',y'))
      \sqcup \\
      &\qquad\bigsqcup_{a'>a, x''\in X}\alpha(x,a',x'')\big),
    \end{align*}
    i.e. we have
    \begin{equation*}\begin{aligned}\phi'\in &\lnot\alpha(x,a,x')\sqcup\big(\bigsqcup_{y'\in
      Y}(\beta(y,a,y')\sqcap R(x',y')) \\&\sqcup\bigsqcup_{a'>a, x''\in
      X}\alpha(x,a',x'')\big)\end{aligned}\end{equation*}
    in the Boolean algebra. Since $R(x',y')$ is a lattice element and
    therefore downward-closed, we can apply
    Lemma~\ref{lem:approximpli} and conclude that
    \begin{equation*}\begin{aligned}\phi' \in \bigg(&\alpha(x,a,x') \rightarrow\big(\bigsqcup_{y'\in
      Y}(\beta(y,a,y')\sqcap R(x',y')) \\&\sqcup\bigsqcup_{a'>a, x''\in
      X}\alpha(x,a',x'')\big)\bigg)\end{aligned}\end{equation*}
    in the lattice, concluding the proof.
	\end{itemize}
\end{IEEEproof}

Hence we can compute the bisimulation via a fixpoint iteration, as
with LaTSs without an ordering on the action labels. Due to the additional
supremum in the fixpoint operator, the matrix notation cannot be used
anymore. However, since the additional supremum term can be
precomputed for each pair of states $x\in X$ or $y\in Y$ and action
$a\in A$, the performance of the algorithm should not be affected in a
significant way.

Note that, different from the Boolean case, $l_1\rightarrow (l_2\sqcup l_3) \not\equiv (l_1\rightarrow l_2)\sqcup l_3$,
which is relevant for the definition of $G$. In fact, moving the
supremum $\bigsqcup_{a'>a, x''\in X}\alpha(x,a',x'')$ outside of the
residuum would yield an incorrect notion of bisimilarity.

In addition, it may appear more convenient to drop the monotonicity
requirement for transitions and to allow arbitrary deactivation of
transitions, independently of their label. However, this would result
in a loss of the duality result and as a result, the fixpoint
algorithm that allows to compute the bisimilarity in parallel for all
products would be rendered incorrect.
}

\end{document}